\documentclass[a4paper,fleqn,usenatbib]{mnras}

\usepackage{newtxtext,newtxmath}

\usepackage[T1]{fontenc}
\usepackage{ae,aecompl}

\usepackage{graphicx}	
\usepackage{amsmath}	
\usepackage{amssymb}	
\usepackage{subcaption}
\usepackage[authoryear]{natbib}
\bibpunct{(}{)}{;}{a}{}{,}
\usepackage{graphicx}
\usepackage{color}
\usepackage{amssymb}
\usepackage{ulem}
\usepackage{natbib}
\usepackage{courier}
\title[Triggering AGN in galaxy clusters]{Triggering Active Galactic Nuclei in galaxy clusters}

\author[M. A. Marshall et al.]{Madeline A. Marshall$^{1}$\thanks{E-mail: madelinem1@student.unimelb.edu.au}, Stanislav S. Shabala$^{1}$, Martin G. H. Krause$^{1,2}$,
\newauthor Kevin A. Pimbblet$^{3,4,5}$, Darren J. Croton$^6$, Matt S. Owers$^{7,8}$
\\
$^{1}$ School of Physical Sciences, Private Bag 37, University of Tasmania, Hobart, TAS 7001, Australia\\
$^{2}$ Centre for Astrophysics Research, University of Hertfordshire, College Lane, Hatfield, Herts AL10 9AB, UK\\
$^{3}$ Milne Centre for Astrophysics, University of Hull, Cottingham Road, Kingston-upon-Hull, HU6 7RX, UK\\
$^{4}$ School of Physics, Monash University, Clayton, VIC 3800, Australia\\
$^{5}$ Monash Centre for Astrophysics (MoCA), Monash University, Clayton, VIC 3800, Australia\\
$^{6}$ Centre for Astrophysics and Supercomputing, Swinburne University of Technology, P.O. Box 218, Hawthorn, VIC 3122, Australia\\
$^{7}$ Department of Physics and Astronomy, Macquarie University, NSW 2109, Australia\\
$^{8}$ Australian Astronomical Observatory, P.O. Box 915, North Ryde, NSW 1670, Australia
}

\date{Accepted XXX. Received YYY; in original form ZZZ}

\pubyear{2017}

\begin{document}
\label{firstpage}
\pagerange{\pageref{firstpage}--\pageref{lastpage}}
\maketitle

\begin{abstract}
We model the triggering of Active Galactic Nuclei (AGN) in galaxy clusters using the semi-analytic galaxy formation model SAGE \citep{Croton16}. We prescribe triggering methods based on the ram pressure galaxies experience as they move throughout the intracluster medium, which is hypothesized to trigger star formation and AGN activity. The clustercentric radius and velocity distribution of the simulated active galaxies produced by these models are compared with that of AGN and galaxies with intense star formation from a sample of low-redshift, relaxed clusters from the Sloan Digital Sky Survey. The ram pressure triggering model that best explains the clustercentric radius and velocity distribution of these observed galaxies has AGN and star formation triggered if $2.5\times10^{-14}<P_{\textrm{ram}}<2.5\times10^{-13}$ Pa and $P_{\textrm{ram}}>2P_{\textrm{internal}}$; this is consistent with expectations from hydrodynamical simulations of ram-pressure induced star formation. Our results show that ram pressure is likely to be an important mechanism for triggering star formation and AGN activity in clusters.
\end{abstract}




\section{Introduction}
\label{sec:intro}


Recent decades have brought recognition of the important role Active Galactic Nuclei (AGN) play in regulating the cosmic star formation histories of their host galaxies and larger scale environments (e.g. \citealt{Bohringer}, \citealt{SilkRees}, \citealt{Croton}, \citealt{Bower}, \citealt{Vogelsberger}; see \citealt{Fabian} and \citealt{HeckmanBest} for reviews).
Whilst a consensus on the existence of a relationship between AGN presence and host galaxy environment is yet to be reached, the majority of the current evidence seems to suggest that the presence of AGN has a strong environmental dependence, with factors such as the local galaxy density and one-on-one interactions highly influencing the likelihood of a galaxy hosting an AGN \citep{Sabater}. AGN are powered by the accretion of gas onto the supermassive black holes that reside at the centres of most galaxies.
For this to occur, there must be an abundant supply of gas in the central regions of the galaxy \citep[e.g.][]{Reichard}. The environmental dependence of AGN activity is a consequence of this requirement, with the availability of gas dependent on environmental effects; for example, ram pressure stripping in the inner regions of clusters may deplete a galaxy's gas supply, whilst galaxy mergers may inject a new supply of gas to the galaxy.

Whilst an abundance of cold gas is a requirement for AGN activity, it is also essential for star formation to occur; since both processes are theoretically expected to rely on a supply of cold gas, the two are expected to be fundamentally linked. Such a correlation between AGN activity and star formation is discussed in the literature and consistent with observations \citep[e.g.][]{Mullaney, Diamond-Stanic,Hickox14}, with AGN tending to exist in hosts with ongoing star formation \citep{Rafferty,BestHeckman} and with similarities observed in the redshift evolution of star formation and black hole accretion \citep[e.g.][]{Shankar,Driver}. In fact, an evolutionary sequence from star-forming galaxies via AGN to quiescence has been proposed \citep{Schawinski}, suggesting that the two processes are caused by the same mechanism but with a substantial delay between the starburst and AGN activity of roughly 250 Myr. Analogous conclusions are made by \citet{Wild} and \citet{Shabala}; also see \citet{Krause05}.

AGN and star formation are triggered by processes which increase the supply of gas to the centres of galaxies. For example, mergers and interactions of galaxies are expected to induce AGN activity, since they provide a torque on the gas which can funnel it into the centre of the galaxy \citep{Hernquist}. 
This is supported by observations, with an observed increase in the prevalence of AGN in interacting galaxy pairs \citep{Sabater,Ellison} and `lopsided' galaxies \citep{Reichard}, which are indicative of galaxy interactions. Observations and simulations show that mergers also trigger episodes of star formation \citep[e.g.][]{Doyon,Mihos}, with an AGN preceded by a starburst a common evolutionary sequence during gas-rich mergers \citep{Hopkins,Melnick}.

Alongside galaxy mergers, the ram pressure that cluster galaxies experience as they move through the intracluster medium (ICM) may induce AGN and star formation.
This pressure is able to strip gas from galaxies (ram pressure stripping), which may cause tails of stripped gas to form behind the galaxy as it moves in the cluster, as has been observed for galaxies in the Virgo cluster \citep[e.g.][]{Kenney,Crowl,Chung}. Ram pressure stripping leads to a decreased prevalence of radiative-mode AGN activity in the centres of clusters \citep{Ellison,Ehlert2,Khabib}, since the ram pressure has depleted the gas supply of these central galaxies. However, models and hydrodynamical simulations show that lower ram pressures can compress the gas in the galaxy, and actually induce star formation \citep[e.g.][]{Fujita99,Kronberger,Tonnesen,Kapferer,Bekki}, which is also supported by observations \citep[e.g.][]{Lee}. These moderate ram pressures could also conceivably lead to higher black hole accretion, and hence trigger AGN activity, since ram pressure can lead to angular momentum loss in gas clouds \citep{Tonnesen} and trigger gravitational instability in the galactic disk \citep{Schulz}, potentially leading to gas being deposited into the galaxy centre. Additional processes such as frequent high-speed galaxy encounters (galaxy harassment; \citealt{Moore}) and tidal interactions between the galaxy and the cluster potential \citep{Byrd} may also induce gas flows to the centre of a galaxy, triggering AGN.

Ram-pressure and merger-induced AGN activity are dependent on the galaxy's location within the cluster. Hence, studies of the location of AGN in clusters are important for determining the mechanism by which they are triggered. \citet{Pimbblet} considered a sample of emission-line AGN in 6 relaxed clusters from the Sloan Digital Sky Survey (SDSS) in $0.070<z<0.089$. They found that the radial AGN fraction increases steeply in the central 1.5 virial radii ($r_{\textrm{vir}}$) of clusters, but flattens off quickly and even decreases beyond this radius. \citet{Ruderman} studied X-ray AGN in 24 virialized clusters in $0.3 < z <0.7$ and found a radial AGN density profile with a pronounced peak in the centre ($r <0.5$ Mpc), with a secondary broad peak at approximately the virial radius (2-3 Mpc). They attribute the central spike to close encounters between infalling galaxies and the giant cD-type elliptical galaxy at the cluster centre, and suggest that the secondary peak at the viral radius is due to an increase of galaxy mergers in these regions. A similar result is found by \citet{Ehlert1} who investigated 43 of the most massive and X-ray luminous clusters in $0.2<z<0.7$; they found an excess of X-ray AGN in the cluster centres, and a secondary excess at around the viral radius. \citet{Pentericci} found that moderately-luminous X-ray AGN are found in the outer regions of clusters at $z\sim0.5$-1.1 relative to normal galaxies, but have the same velocity distribution.

\citet{Pimbblet} found that the most powerful AGN reside in the cluster infall regions. \citet{Haines} also reach the same conclusion for X-ray and optical AGN; these were found to have a higher velocity dispersion in comparison to the general cluster population and avoid the phase space with low relative velocities and clustercentric radii. In addition, \citet{Martini02} found a higher mean velocity offset for X-ray AGN in a massive cluster A2104 ($z=0.154$) than for other cluster members, suggesting that some of the AGN may be falling into the cluster.
These studies suggest that AGN are triggered by some mechanism which acts on galaxies during their infall, producing AGN around the virial radius of the cluster; this provides a vital clue as to how cluster AGN are triggered. Comparing expectations of the various hypothesized triggering methods to such observational samples may give an important insight into the triggering mechanisms of AGN in clusters. Similar analyses for star-forming galaxies are more complicated, due to the difficulty in separating induced and secular star formation activity.

In this paper, we test the relationship between AGN activity and the location of cluster galaxies in the position-velocity phase space. In particular, we test the hypothesis that ram pressure effects on satellite (i.e. not cluster-central) galaxies can trigger galaxy activity, by simulating ram pressure triggering on model galaxies and comparing the spatial distribution of the resulting active galaxies to observations. This paper is organised as follows. In Section \ref{sec:model}, we outline the semi-analytic model used to simulate the evolution of cluster galaxies, and detail the method used for triggering AGN. The model is validated in Section \ref{sec:results} with a comparison to the \citet{Pimbblet} observations, with predictions of the model and a comparison with additional datasets given in Section \ref{sec:ModelPredictions}. The implications of these results are discussed in Section \ref{sec:discussion}, before concluding in Section \ref{sec:conclusions}.
Throughout this paper we assume $h=0.73$, $\Omega_m = 0.25$ and $\Omega_\Lambda = 0.75$.

\section{AGN triggering in a semi-analytic galaxy formation model}
\label{sec:model}

\subsection{Model galaxies}
\subsubsection{Semi-Analytic Galaxy Evolution model}
The Semi-Analytic Galaxy Evolution model, or SAGE, originally introduced in \citet{Croton} and updated in \citet{Croton16}, models galaxy formation and evolution in a cosmological context. 
SAGE analytically models the baryonic physics involved in galaxy formation and evolution, such as gas infall and cooling, star formation, black hole growth, AGN and supernova feedback, and reionization. To do so, semi-analytic models are implemented on the output of an N-body cosmological simulation of dark matter, with the baryons added in post-processing. The \citet{Croton16} version of SAGE can be run on a number of N-body simulations, however we implement the model based on the Millennium Simulation \citep{Springel}. This simulation contains $2160^3$ particles with a particle mass of $8.6\times10^8 h^{-1}\textrm{M}_\odot$ in a periodic box of $(500h^{-1}\textrm{Mpc})^3$. Dark matter halos within the simulation are identified using halo-finding algorithms, with particles with an overdensity of around 200 -- the overdensity roughly expected of a virialized group -- grouped together, and traced forward in time. Baryons are then planted on these halos and subhalos and their evolution followed by the galaxy model prescriptions. The reader is referred to \citet{Croton16} for a full description of SAGE.

SAGE uses the properties of each dark matter halo and subhalo in the simulation at 64 discrete redshift snapshots from $z=127$ to $z=0$. We consider only the lowest redshift snapshots for the majority of our analysis. Properties of each galaxy given by the semi-analytic model at each snapshot include its total stellar mass, the stellar mass of the bulge, black hole mass, cold gas mass, hot gas mass and disk scale radius, alongside its position and velocity in the three dimensions of the simulation box. The properties of each cluster/dark matter halo include its virial mass and radius, with an identification system such that all cluster members can be easily identified. Galaxies can be followed from one redshift slice to another using their unique galaxy IDs.

\subsubsection{Post-processing}
The density profile of the ICM for each cluster is approximated using cluster density profiles determined from Chandra X-ray observations \citep{Fujita,Vikhlinin}. If the virial mass and radius of the simulated cluster are within 20\% of those of clusters observed by \citet{Fujita} or \citet{Vikhlinin}, the observed density profile for the closest matching cluster is used. Otherwise, the cluster is given the average density profile of the \citet{Vikhlinin} clusters with $T>2.5$ keV or $T<2.5$ keV, dependent on the virial temperature of the simulated cluster. From the interpolated SAGE positions and velocities, the ram pressure $P_{\textrm{ram}}=\rho_{\textrm{ICM}} v^2$ for each galaxy is calculated using its velocity with respect to the cluster and the ICM density at its distance from the cluster centre, as given from the assumed density profile.

The internal pressure of each cluster galaxy is calculated by assuming pressure equilibrium with the surrounding ICM, which is taken to be at the virial temperature. These thermal pressures depend only on the galaxy's distance from the cluster centre and the properties of the cluster, in contrast to the commonly implemented \citet{Blitz} scaling relation, which instead considers the galaxy's stellar mass, gas mass and radius. The thermal pressures for the galaxies considered here, which span a range of roughly 5 orders of magnitude, are on average $1.2\substack{+0.9 \\ -1.1}$ dex smaller than the pressures calculated using the \citet{Blitz} scaling relation with parameters as typically observed in galaxies; the \citet{Blitz} relation relies on information such as the disk scale height, gas velocity dispersion and stellar and gas disk surface density profiles, that are unknown for our simulated galaxies. The larger \citet{Blitz} internal pressures give lower ratios of the ram pressure to the internal pressure for each galaxy; the threshold pressure ratio for AGN triggering that we determine would be lower if we used these internal pressures, however the absolute ram pressure triggering range we determine would be unaffected.

\subsection{AGN triggering}
In this section, we describe our prescriptions for AGN triggering.

\subsubsection{Ram pressure}
\label{sec:rampressure}
\begin{figure}
\begin{center}
\includegraphics[scale=0.65]{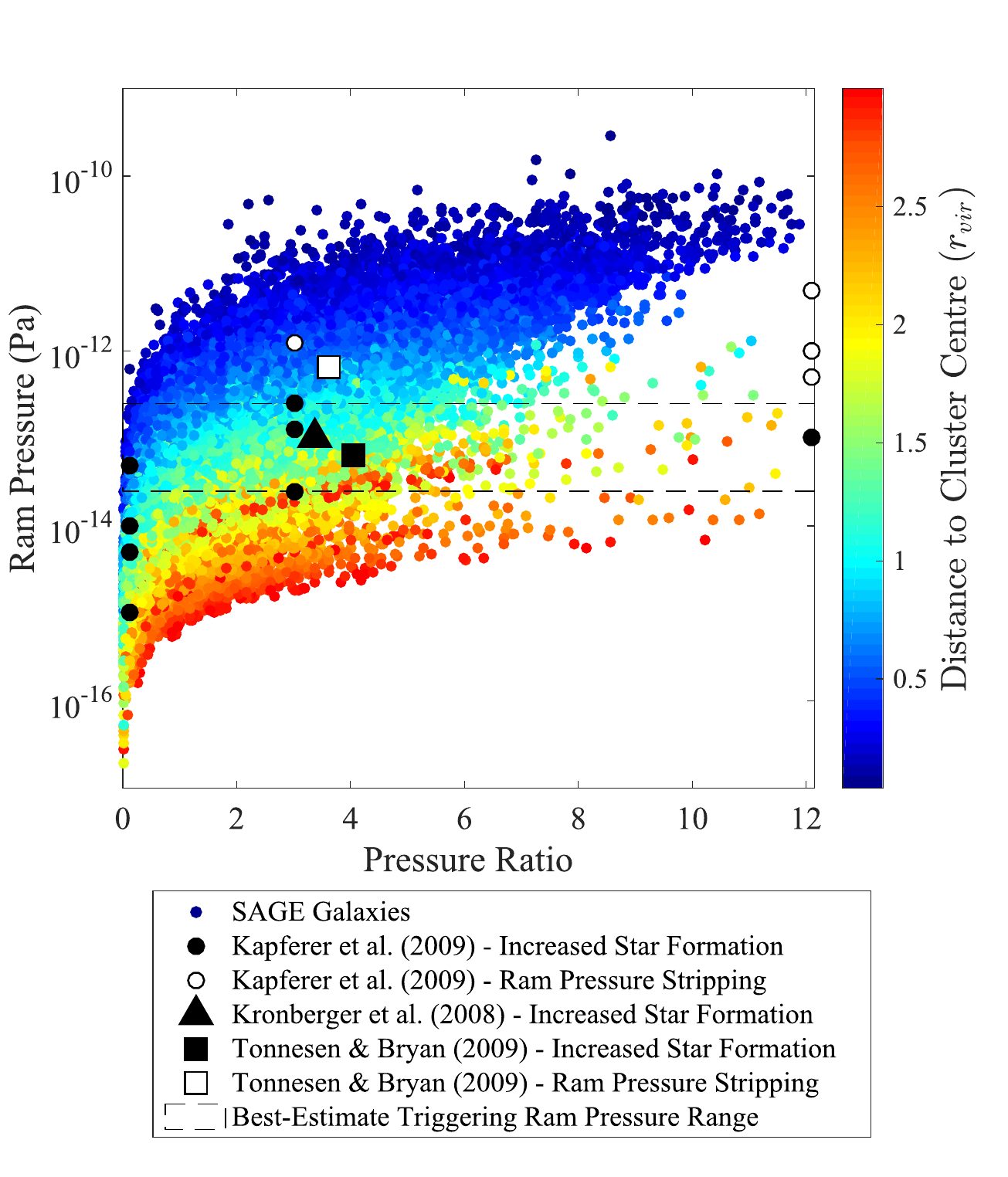} 
\caption{Ram pressure versus pressure ratio for each of the \citet{Kronberger}, \citet{Kapferer} and \citet{Tonnesen} simulations and the semi-analytic galaxies in our SAGE sample. SAGE galaxies are colour-coded by their distance from the cluster centre. Labelled are the simulations which found an increase in star formation (black) and those that resulted in ram pressure stripping (white). From these simulations we determine a best-estimate range of triggering ram pressures of $2.5\times10^{-14}\lesssim P_{\textrm{ram}}\lesssim 2.5\times10^{-13}$ Pa (outlined by dashed lines), corresponding to the medium triggering pressure range considered in our model. Note that the size of the points for the \citet{Kapferer} galaxies are smaller than the symbols for the \citet{Kronberger} and \citet{Tonnesen} galaxies; this is to reflect that the \citet{Kapferer} simulated galaxies are $>0.5$ dex less massive than the other simulated galaxies. With a mass of $2\times10^{10}\textrm{M}_\odot$, these simulated galaxies are in fact below the mass cut used by \citet{Pimbblet}, which we adopt for our SAGE galaxies here, and hence we only show these simulated galaxies here for comparison.}
\label{SimulationPressures}
\end{center} 
\end{figure}

We trigger AGN based on the ram pressure acting upon each galaxy. Hydrodynamical simulations such as those by \citet{Kronberger}, \citet{Kapferer} and \citet{Tonnesen} have shown that ram pressure can cause an increase in star formation in a galaxy, or can strip the gas from the galaxy, depending on the amount of ram pressure the galaxy experiences. We assume that ram pressures which cause an increase in star formation can trigger AGN activity. The ram pressures and ratios of ram pressure to internal pressure of galaxies considered in these simulations are shown in Fig. \ref{SimulationPressures}, with simulations resulting in increased star formation and those which cause ram pressure stripping distinguished. Fig. \ref{SimulationPressures} clearly shows that above ram pressures of approximately $2.5\times10^{-13}$ Pa, galaxies undergo ram pressure stripping, whilst below this critical value the star formation in the galaxy can be increased by the ram pressure. Note that simulated SAGE galaxies with low clustercentric radii are less frequently found at ram pressures below $10^{-14}$ Pa, with those that exist in the simulation having low pressure ratios; galaxies with these low ram pressures are unlikely to have ram-pressure induced star formation.
Therefore, taking a best-estimate range of triggering ram pressures that spans an order of magnitude, we expect that ram pressure is most likely to cause an increase in star formation and therefore AGN activity if
$2.5\times10^{-14}\lesssim P_{\textrm{ram}}\lesssim 2.5\times10^{-13}$ Pa. Guided by this expectation, we explore three triggering ram pressure ranges: $2.5\times10^{-15}<P_{\textrm{ram}}<2.5\times10^{-14}$ Pa (Low), $2.5\times10^{-14}<P_{\textrm{ram}}<2.5\times10^{-13}$ Pa (Medium), and $2.5\times10^{-13}<P_{\textrm{ram}}<2.5\times10^{-12}$ Pa (High).\\

\subsubsection{Pressure ratio}

In addition to the variations of the triggering ram pressures, we also consider the ratio of the ram pressure to the internal pressure of the galaxy when triggering the simulated AGN.
Figure \ref{SimulationPressures} shows that the pressure ratio is unimportant when determining whether a galaxy undergoes ram pressure stripping in the hydrodynamic simulations. Nevertheless, our physical intuition tells us that the ram pressure must at least be comparable to the internal pressure in the galaxy to have an effect. 
If the ram pressure is small relative to the internal pressure, then the galaxy would not be expected to undergo a significant change in response to the ram pressure, even if its magnitude is in the triggering range.
We therefore consider lower limits on the pressure ratios that can trigger AGN, heuristically chosen to be $\frac{P_{\textrm{ram}}}{P_{\textrm{internal}}}=0$, 1 or 2, in the absence of evidence for strong differences in this regard within the parameter range explored.

\subsubsection{AGN luminosity model and time delay}
\label{sec:timedelay}
The luminosities of the triggered AGN are not directly attainable from the simulations without making significant assumptions. To estimate the simulated AGN luminosities, we assume that the ratio of bolometric to Eddington luminosity $L_{\textrm{Bol}}/L_{\textrm{Edd}}=0.1$\footnote{A ratio of $L_{\textrm{Bol}}/L_{\textrm{Edd}}=0.1$ is reasonable for radiative-mode AGN, and is within 1$\sigma$ of the median $L_{\textrm{Bol}}/L_{\textrm{Edd}}$ fraction for the observed sample.} and use the known black hole mass to convert this to a value in $L_\odot$. 

Studies such as \citet{Schawinski}, \citet{Wild}, \citet{Shabala}, \citet{Krause05} and \citet{Melnick} suggest that star formation and AGN activity may be caused by a common mechanism, with a substantial delay between the starburst and the onset of the AGN activity, of roughly 250 Myr.
We investigate the effect of such a time delay on the resulting AGN distribution. We consider the case of no time delay, in which AGN are triggered instantaneously if the ram pressure satisfies the triggering conditions, and a time delay of 250 Myr, in which the galaxy's properties 250 Myr ago are used to determine whether an AGN is triggered at $z=0$. Since no SAGE snapshot at a lookback time of 250 Myr ago exists, to determine the galaxy properties at this time we consider the 5 lowest redshift snapshots, at $z=0.000$, 0.020, 0.041, 0.064 and 0.089 (corresponding to lookback times of 0, 277, 560, 860 and 1175 Myr, respectively), and interpolate the positions and velocities in intervals of approximately 3 Myr.

\section{Model validation}
\label{sec:results}
\subsection{Validation sample}
\label{sec:observations}

\subsubsection{The \citet{Pimbblet} sample}
To assess the validity of ram pressure triggering of AGN and star formation, the simulated active galaxies are compared with AGN and galaxies with intense star formation from the \citet{Pimbblet} observational sample. This is a sample of six clusters in $0.070<z<0.089$ observed in the SDSS that have no observable signs of merging or significant interactions with other clusters or subclusters. Such interactions may cause a local enhancement in AGN activity, complicating studies of the effects of the general cluster environment on AGN activity; the \citet{Pimbblet} sample of relaxed clusters is therefore ideal for this work (see Section \ref{sec:complications}).
Of the galaxies targeted by the SDSS in these clusters, only galaxies with $\log{\frac{M_\ast}{M_\odot}}>10.4$ and brighter than $M_r=-19.96$ are considered for completeness. 

The \citet{Pimbblet} study considered only AGN and not star-forming galaxies. \citet{Pimbblet} select emission-line AGN using the commonly used BPT diagnostic \citep{BPT}; these use the ratios of [NII]$_{\lambda6583}$/H$\alpha$ and [OIII]$_{\lambda5007}$/H$\beta$ to differentiate galaxies in which photoionization is caused by hot O and B stars (star-forming galaxies) and those in which photoionization is caused by a non-thermal source, for example, AGN activity \citep{Veilleux}. To be selected as an AGN, \citet{Pimbblet} required galaxies to have a signal to noise ratio (SNR) of greater than 3 in each of the [OIII]$_{\lambda5007}$, [NII]$_{\lambda6583}$, H$\alpha$ and H$\beta$ lines, and lie above the \citet{Kauffmann} demarcation curve -- an empirically determined classification for distinguishing AGN from star-forming galaxies. Galaxies above this curve are classified into three types dependent on their location in the BPT plane: Seyferts, low-ionization nuclear emission-line regions (LINERs) and transition objects (which lie between the \citealt{Kauffmann} and \citealt{Kewley} demarcation curves on the BPT diagram), all of which are included in the \citet{Pimbblet} AGN sample.

We also consider galaxies in the \citet{Pimbblet} sample which have star formation that is significantly greater than typical secular star formation, as given by the star formation main sequence -- `intense star-formers' (Section \ref{sec:SF}). We consider star-forming galaxies because we hypothesise that ram pressure can trigger both star formation and AGN activity, since the two processes are closely linked.

\subsubsection{Observational selection of AGN}
\label{sec:AGNselection}
In order to be confident in the emission-line AGN sample selected from this data, alongside the \citet{Kauffmann} and signal to noise criteria implemented by \citet{Pimbblet} we impose additional, stricter criteria for AGN selection. 

LINERs have ionization signatures that can be produced by low-level AGN activity, however, these can also be produced by cooling flows or shock-heated gas \citep[e.g.][]{Heckman87, Kauffmann}, or by galaxies which have stopped forming stars, with the ionization produced by hot post-AGB stars and white dwarfs \citep{Stasinska}. Therefore, LINERs are not a pure AGN population, unlike Seyfert galaxies.
In order to distinguish the population of LINERs that host AGN from those galaxies which have stopped forming stars, or `retired galaxies', only galaxies with H$\alpha$ equivalent widths ($W_{\textrm{H}\alpha}$) of magnitude greater than 3\AA\ are selected as AGN; this is the condition for non-retired galaxies prescribed by the $W_{\textrm{H}\alpha}$ versus [NII]/H$\alpha$ (WHAN) diagnostic \citep{CidFernandes2011}. 

\begin{figure}
\begin{center}
\includegraphics[scale=0.5]{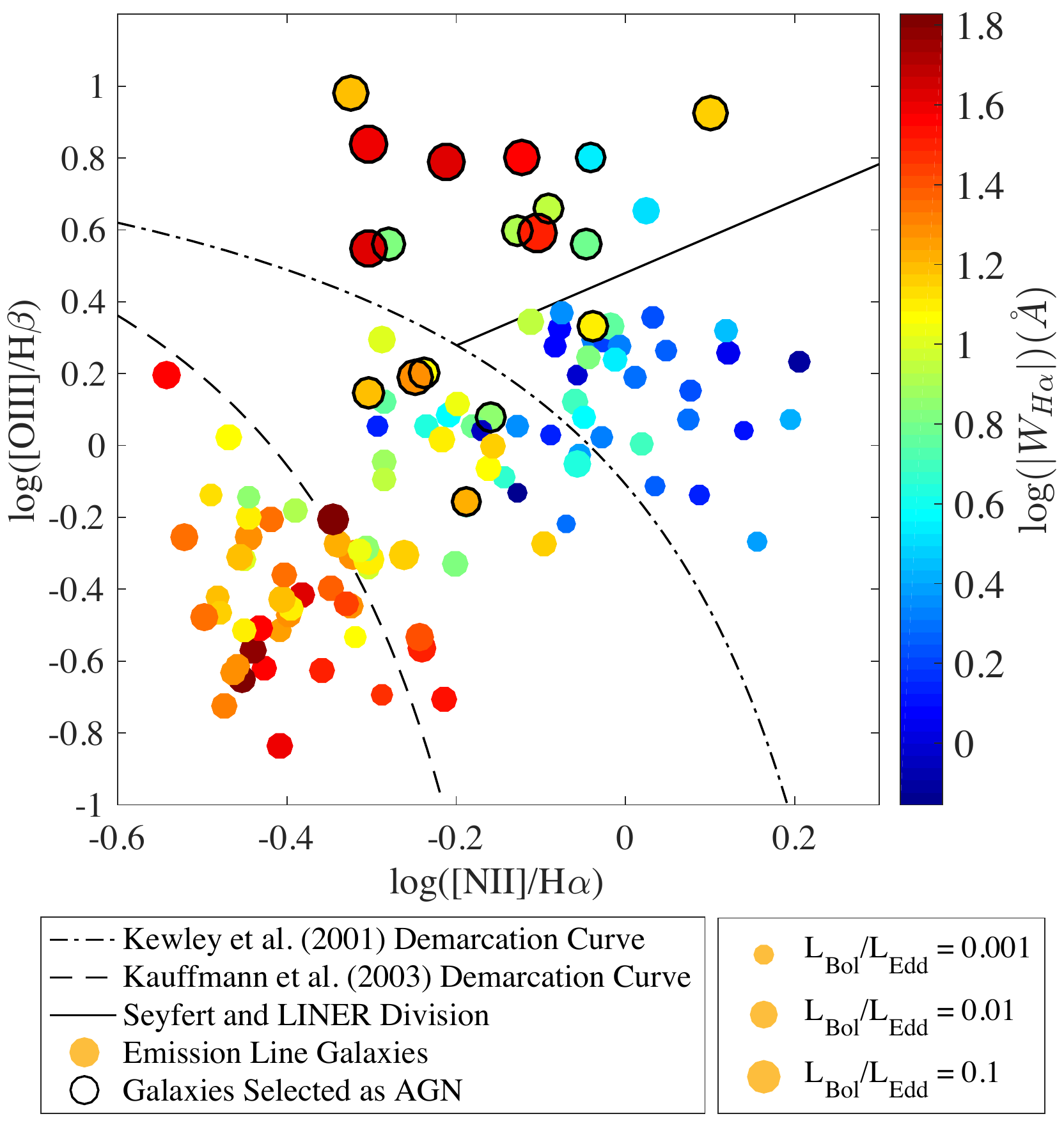} 
\caption{BPT diagram for the \citet{Pimbblet} galaxies with $\textrm{SNR}>3$ in each of the four emission lines and with mass greater than $10^{10.4}\textrm{M}_\odot$. Point colours represent H$\alpha$ equivalent widths and sizes represent $\log{\left(L_{\textrm{Bol}}/L_{\textrm{Edd}}\right)}$, with the largest points corresponding to the largest $L_{\textrm{Bol}}/L_{\textrm{Edd}}$ values. AGN (black circles) are those galaxies which satisfy: $|W_{\textrm{H}\alpha}|>3$\AA\ (yellow-green points), $L_{\textrm{Bol}}/L_{\textrm{Edd}}>0.01$ (large points), lie above the \citet{Kauffmann} demarcation curve (dashed curve) and are closer to the \citet{Kewley} demarcation curve (dot-dashed curve) than to the \citet{Kauffmann} curve.}
\label{BPT}
\end{center} 
\end{figure}

Studies have found a dichotomy in the accretion rates of radiative-mode and jet-mode radio-loud AGN relative to the Eddington rate; radiative-mode AGN accrete at higher rates, with bolometric luminosities of 0.01--1 of the Eddington luminosity, whilst jet-mode AGN have lower bolometric luminosities of less than 0.01 of the Eddington luminosity \citep[see e.g.][]{BestHeckman,Daly}. Whilst radiative-mode AGN are expected to be triggered by interactions of the galaxy with its local environment, jet-mode AGN are more often triggered by large-scale cooling of gas onto the galaxy \citep{Best,Shabala08,Fabian}; only radiative-mode AGN are of interest in our study.
Therefore, motivated by this dichotomy of accretion rates, we consider only galaxies with ratio of bolometric to Eddington luminosity $L_{\textrm{Bol}}/L_{\textrm{Edd}}>0.01$\footnote{The black hole mass used to calculate the Eddington luminosity is estimated via a black hole -- bulge mass relation \citep{Haring}, with the bulge mass estimated using the galaxy's total stellar mass and its bulge fraction and then converted to a black hole mass.}; this removes jet-mode AGN from the AGN sample. Fig. \ref{BPT} shows the variation of H$\alpha$ equivalent widths and $L_{\textrm{Bol}}/L_{\textrm{Edd}}$ ratios of the \citet{Pimbblet} sample across the BPT diagram.

The resulting AGN sample contains 7 transition objects that lie significantly closer to the \citet{Kauffmann} demarcation curve than to the \citet{Kewley} demarcation curve. Since these are likely to be intense star-formers and not AGN, we exclude these galaxies from the AGN sample. Four of these transition objects are classified as intense star-formers as defined in Section \ref{sec:SF}. Including the additional 3 transition objects in the composite AGN/star-former sample has no effect on the results.
Our final AGN sample contains 18 galaxies, shown on the BPT diagram in Fig. \ref{BPT}.
The galaxies hosting these AGN have observed [OIII] luminosities ranging from $10^{6.3}$--$10^{8.6}L_\odot$, corresponding to bolometric luminosities of roughly $10^{10}$--$10^{12}L_\odot$ using the conversion of \citet{Heckman04}.
We do not attempt to match this observed luminosity distribution when selecting the simulated AGN sample, however, the AGN luminosities of the two samples are consistent: the luminosities of the simulated AGN from the best-fitting ram pressure model (see Section \ref{sec:BestModel}) range from $10^{9}$--$10^{12}L_\odot$, with the majority between $10^{10}$--$10^{12}L_\odot$.

\subsubsection{Selection of galaxies with intense star formation}
\label{sec:SF}
In addition to selecting a sample of AGN, we also consider galaxies with intense star formation. A connection between starburst galaxies and AGN is theoretically expected, with hydrodynamical simulations showing that gas inflows which produce a burst of star formation can also fuel the central black hole to power an AGN \citep[e.g.][]{DiMatteo,Hopkins}. Observations investigating the relation between a galaxy's recent star formation history and AGN activity support this hypothesis \citep[e.g.][]{Wild07}; in other words, intense star-formers and AGN can be caused by the same mechanism. 

\begin{figure}
\begin{center}
\includegraphics[scale=0.55]{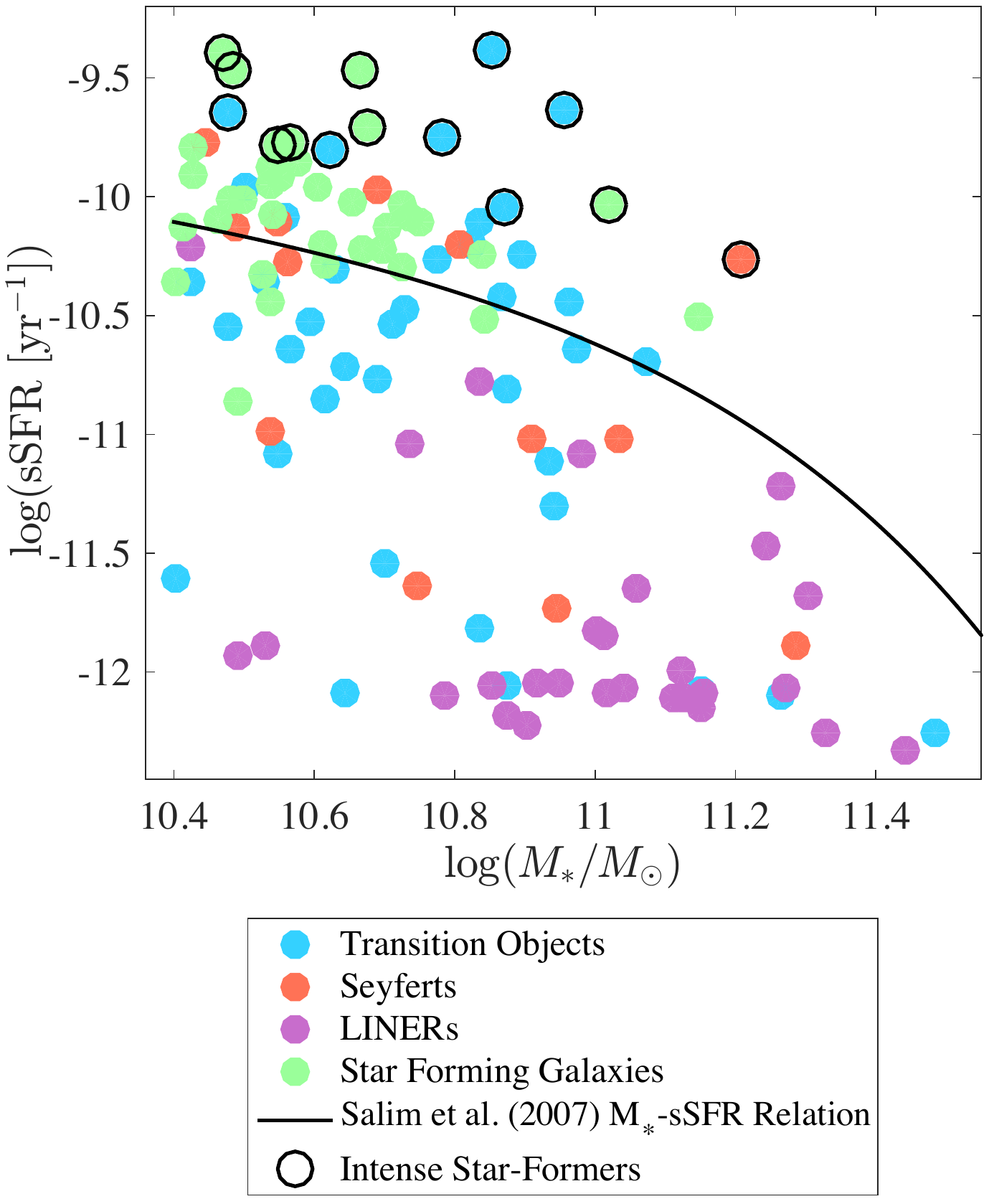} 
\caption{Specific star formation rate (sSFR) against stellar mass for the \citet{Pimbblet} galaxies with $\textrm{SNR}>3$ in each of the four emission lines and with mass greater than $10^{10.4}\textrm{M}_\odot$. Point colours show the BPT classification of each galaxy. Intense star-formers (black circles) have sSFRs above the expected sSFR (black curve) by $>1.5\sigma$.}
\label{sSFR}
\end{center} 
\end{figure}

We consider the $M_\ast$ -- specific star formation rate (sSFR) relation determined from a sample of $\sim50,000$ optically selected galaxies in the local universe ($z\approx0.1$) which range from gas-rich dwarfs to massive ellipticals:
\begin{equation}
\textrm{sSFR}=\textrm{sSFR}_0 \left(\frac{M_*}{M_0}\right)^{\alpha+1}\exp\left(-\left(\frac{M_*}{M_0}\right)^{\alpha+1}\right)
\end{equation}
where $\textrm{sSFR}_0=5.96\times10^{-11}\ \textrm{yr}^{-1}$, $\log M_0=11.03$ and $\alpha=-1.35$ \citep{Salim}, as shown in Fig. \ref{sSFR}.
We define intense star-formers as those that have sSFRs above the expected sSFR by greater than 1.5$\sigma$, where $\sigma$ is the standard deviation of the sSFR about the expected value for galaxies in the sample; Fig. \ref{sSFR} shows these galaxies in the $M_\ast$ -- sSFR plane relative to all other star-formers. 
This results in a sample of 13 star-forming galaxies that are not included in the AGN sample, and one which is also classified as an AGN.
Using a cut of 2$\sigma$ produces results that are not qualitatively different, with the sample size reduced.

\subsubsection{Interactions and substructure identification}
In our model, we consider AGN and star formation which are triggered due to an interaction with the ICM. Hence, for our purposes, observed AGN and intense star-formers that may be explained by galaxy interactions are not relevant. Active galaxies that are part of substructure may be caused by local effects due to the proximity of galaxies within the substructure and increased gas density, with gravitational interactions, mergers and ram pressure stripping more common; this is discussed further in Section \ref{sec:complications}. Hence, active galaxies in substructure are not relevant for the present study.
We therefore consider each of the galaxies in the AGN and intense star-former samples to determine whether they are undergoing interactions or are incorporated in substructure within the cluster.

To identify interacting galaxies, SDSS images of each AGN and star-forming galaxy are examined. 
Galaxies that appear significantly disturbed and have a clear interacting partner (e.g. tidal tails, shells or bridges are clearly identifiable) are classified as `pair' galaxies, with AGN activity or star formation attributed to the interaction. Galaxies that appear significantly disturbed but have no detected nearby companion are likely to be post-mergers, and are also included in this `interacting' or `pair' galaxy classification.

\begin{figure*}
\begin{center}
\includegraphics[scale=0.5]{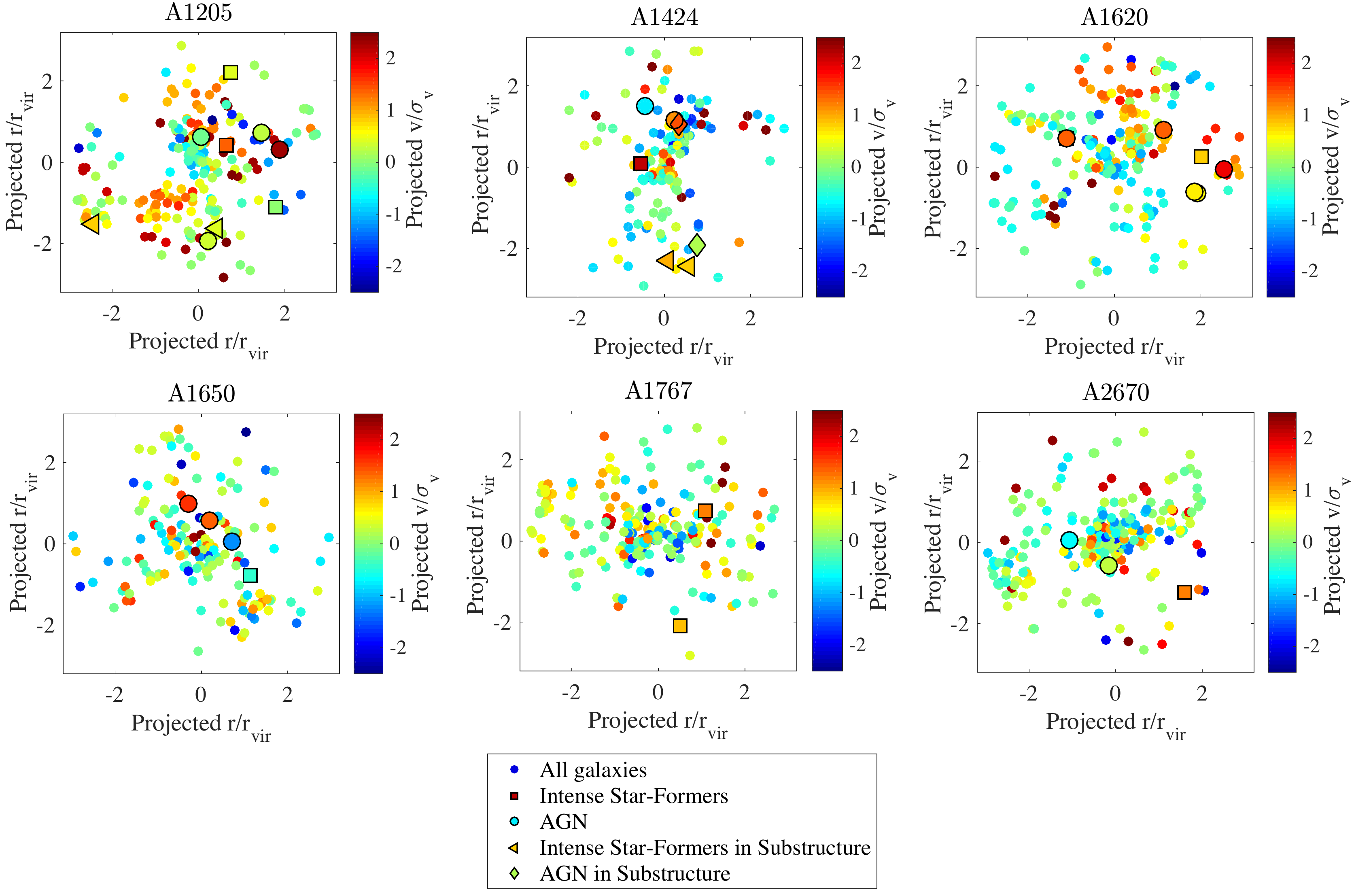} 
\caption{The spatial distribution of the observed galaxies in each of the 6 clusters considered by \citet{Pimbblet}. The points are coloured by the velocity of each galaxy, enabling substructures of galaxies with coherent velocities to be visually identified. Galaxies from the AGN and intense star-former samples that are identified as belonging to substructure by the test discussed in text are clearly distinguished.}
\label{Substructure}
\end{center} 
\end{figure*}

To identify substructure, we search for galaxies around which there is an overdensity of galaxies with a coherent velocity structure (see Fig. \ref{Substructure}). To quantify the level of substructure, first the velocity dispersion of each AGN/star-forming galaxy and its 10 nearest neighbours is calculated. 
We then calculate the velocity dispersion of the AGN/star-forming galaxy and 10 other galaxies in the cluster which lie within the same range of clustercentric radii as the AGN/star-forming galaxy and its 10 nearest neighbours. This is repeated for 1000 groups of 10 galaxies to give a measure of the median and typical spread in the velocity dispersions of galaxies at those clustercentric radii. We then compare the velocity dispersion of the AGN/star-forming galaxy and its 10 nearest neighbours to the median measure. If the velocity dispersion of the neighbours is significantly ($>2\sigma$) less than the median, then the AGN/star-forming galaxy is classified as part of a substructure. 

Of the 31 AGN/star-forming galaxies in the original sample, 10 are identified as `interacting' or `pair' galaxies and 6 are classified as part of a substructure, including 3 within both categories.
Removing AGN/star-forming galaxies affected by interactions or substructure effects results in a final AGN/star-forming galaxy sample containing 18 galaxies. Their locations in phase space are shown in Fig. \ref{PimbbletPP}.
  
\subsubsection{Matching the simulated sample to the observations}
The sample of SAGE galaxies considered in this study is chosen to match the \citet{Pimbblet} observational sample. We select SAGE clusters with virial radii between 20\% less than the smallest and 20\% larger than the largest of the \citet{Pimbblet} clusters, i.e. those with $1.25<r_{\textrm{vir}}/\textrm{Mpc}<2.41$.
This results in a sample containing clusters with $2.2\times10^{14}<M_{\textrm{vir}}/\textrm{M}_\odot<1.6\times10^{15}$ and $500<\sigma_v/\textrm{km\ s}^{-1}<1150$.
In addition, only galaxies in these clusters with $M_\ast>10^{10.4}\textrm{M}_\odot$ are considered, to match the mass cut implemented by \citet{Pimbblet} to ensure completeness. Finally, only galaxies at radii less than $3r_{\textrm{vir}}$ are considered; this excludes galaxies in the cluster outskirts so that all satellites are robustly associated with the cluster.
This results in a sample containing 33931 galaxies from 963 clusters.

We calculate the projected locations and velocities of our simulated galaxies, in order to draw direct comparisons with observational data.
In the simulation cube, galaxies are viewed in projection along each of the $x$, $y$ and $z$ axes.
 
To address selection effects present in the observational dataset, we match the distributions of mass and projected clustercentric radii of the simulated galaxies to that of the \citet{Pimbblet} galaxies. This results in significantly fewer simulated galaxies, specifically at low radii, since the simulated galaxies have a distribution of clustercentric radii that is peaked at small radii and drops significantly, whilst the distribution of clustercentric radii of the observed galaxies is relatively flat, likely at least in part due to SDSS fibre collisions in the central regions of clusters.

\subsection{Location of AGN and star-forming galaxies}
In the following sections, all analysis is completed with the $z=0$ snapshot of SAGE, unless otherwise specified.
\subsubsection{Ram pressure models}
\label{sec:BestModel}
\begin{figure*}
\begin{center}
\includegraphics[scale=0.6]{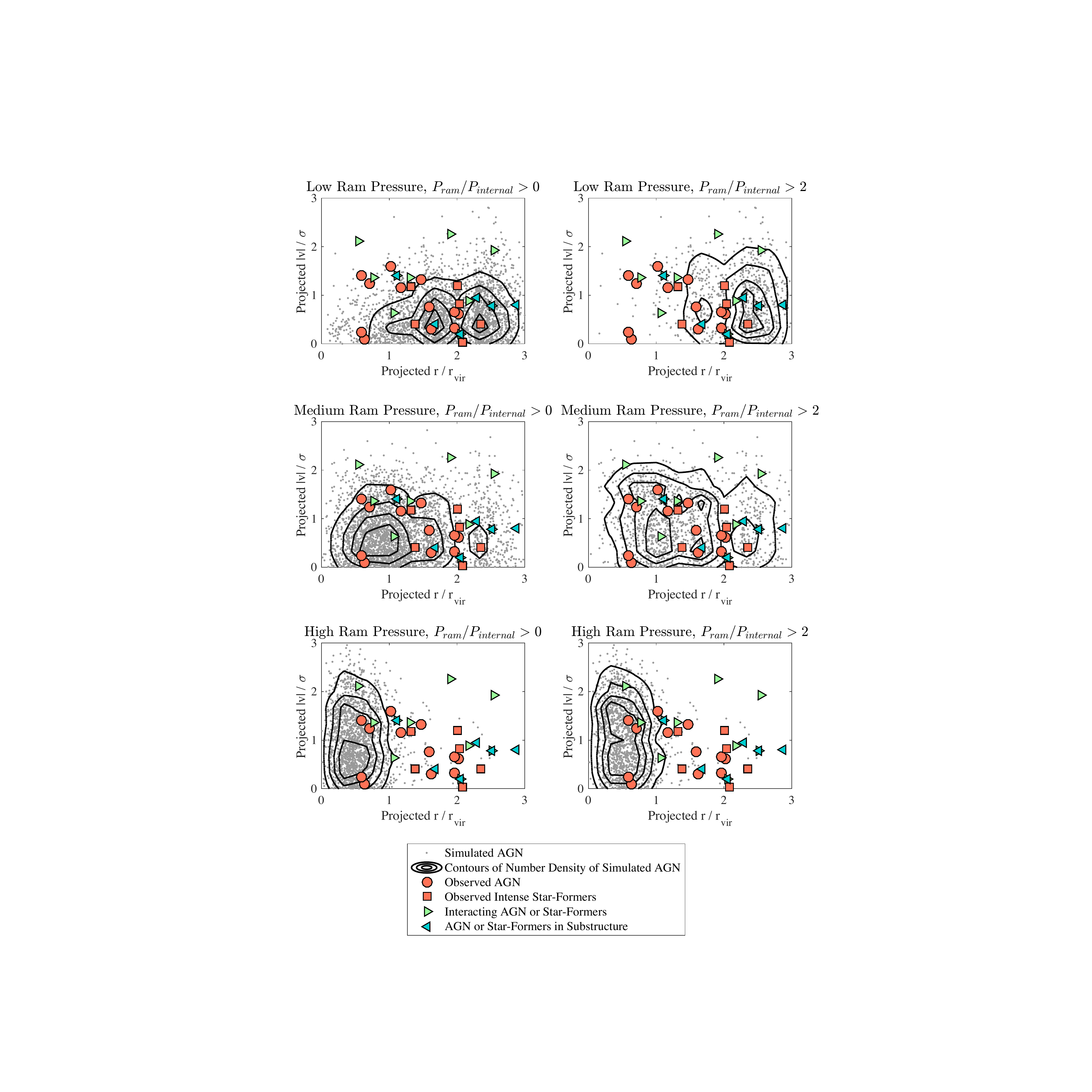} 
\caption{Phase-space diagrams of the simulated AGN produced under various ram pressure triggering models, with the simulated sample matched in radial and mass distributions to the \citet{Pimbblet} sample.
The left column shows changing ram pressure threshold, with constant pressure ratio threshold of 0. The right column shows changing ram pressure threshold, with constant pressure ratio threshold of 2.
Triggering ram pressure ranges are $2.5\times10^{-15}<P_{\textrm{ram}}<2.5\times10^{-14}$ Pa (Low), $2.5\times10^{-14}<P_{\textrm{ram}}<2.5\times10^{-13}$ Pa (Medium), and $2.5\times10^{-13}<P_{\textrm{ram}}<2.5\times10^{-12}$ Pa (High). Contours show the number density of simulated AGN at levels of 20, 40, 60 and 80\% of the peak number density, with the individual AGN also plotted. Superimposed are the AGN and intense star-formers from the observational sample, with galaxies associated with interactions or substructure indicated. The model which best describes the observations is the $2.5\times10^{-14}<P_{\textrm{ram}}<2.5\times10^{-13}$ Pa, $P_{\textrm{ram}}/P_{\textrm{internal}}>2$ model (middle right panel). Lower triggering ram pressures (top panels) produce AGN at larger radii, whilst higher triggering ram pressures (bottom panels) produce AGN at clustercentric radii that are too small to match observations. In comparison to the $P_{\textrm{ram}}/P_{\textrm{internal}}>0$ models (left panels), the $P_{\textrm{ram}}/P_{\textrm{internal}}>2$ models (right panels) lead to less AGN at small radii for a constant triggering ram pressure range.}
\label{PimbbletPP}
\end{center} 
\end{figure*}

We trigger our simulated galaxies as AGN under multiple triggering ram pressure and pressure ratio conditions, with the phase-space distributions of AGN produced by various combinations of these shown in Fig. \ref{PimbbletPP}. 
These are without the effect of a time delay between the onset of enhanced star formation and AGN activity, which is discussed in detail in Section \ref{sec:timedelayresults} below.
These simulated galaxies have radial and mass distributions matched with the \citet{Pimbblet} sample.

The simulated AGN in the model with triggering ram pressures of $2.5\times10^{-14}<P_{\textrm{ram}}<2.5\times10^{-13}$ Pa and $P_{\textrm{ram}}/P_{\textrm{internal}}>2$ are clearly seen to fit the observations best (Fig. \ref{PimbbletPP}, middle right panel). This best-fitting ram pressure range is consistent with expectations from hydrodynamical simulations discussed in Section \ref{sec:rampressure}. Lower triggering ram pressures give AGN at clustercentric radii that are too large, regardless of the triggering pressure ratios, whilst the opposite is true for high triggering ram pressures, with AGN produced at low clustercentric radii between 0--1$r_{\textrm{vir}}$. Adjusting the model to allow galaxies with lower pressure ratios to become AGN results in more AGN at low radii, regardless of the triggering ram pressures.

The changes in velocity distribution with triggering model are secondary to the changes in the radial distribution. For lower pressure ratio thresholds and fixed triggering ram pressures, the AGN are found at lower projected velocities. The AGN are also found at lower projected velocities for models with lower triggering ram pressures, with a fixed pressure ratio threshold. 
All models have a velocity distribution that is roughly consistent with the observations.

\begin{figure}
\begin{center}
\includegraphics[scale=0.6]{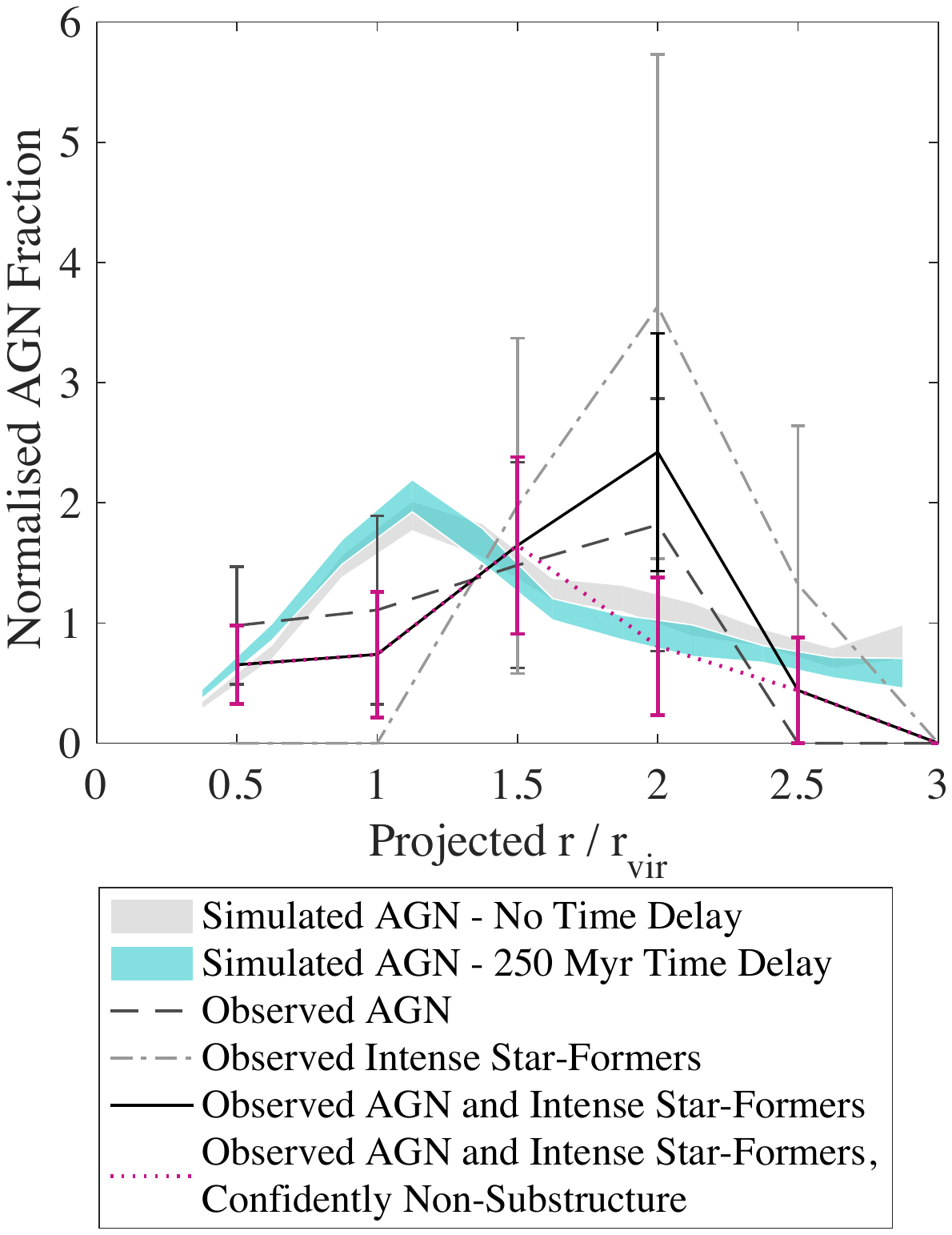} 
\caption{The AGN fraction as a function of clustercentric radius for the simulated AGN from the best ram pressure model (middle right panel of Fig. \ref{PimbbletPP}), alongside the fraction of AGN and intense star-formers (both individually and combined) in the observed sample, with the AGN fractions normalised to the overall fraction of the sample. 
The fraction of simulated AGN from models with no time delay and a 250 Myr time delay are both shown, with no significant differences found between the two. Radial bins for the observational sample are 0-0.75$r_{\textrm{vir}}$, 0.75-1.25$r_{\textrm{vir}}$, 1.25-1.75$r_{\textrm{vir}}$, 1.75-2.25$r_{\textrm{vir}}$, 2.25-2.75$r_{\textrm{vir}}$ and $>$2.75$r_{\textrm{vir}}$, whilst those for the simulated sample are finer: 0-0.5$r_{\textrm{vir}}$, 0.5-0.75$r_{\textrm{vir}}$, 0.75-1$r_{\textrm{vir}}$, 1-1.25$r_{\textrm{vir}}$, 1.25-1.5$r_{\textrm{vir}}$, 1.5-1.75$r_{\textrm{vir}}$, 1.75-2$r_{\textrm{vir}}$, 2-2.25$r_{\textrm{vir}}$, 2.25-2.5$r_{\textrm{vir}}$, 2.5-2.75$r_{\textrm{vir}}$ and $>$2.75$r_{\textrm{vir}}$. Error bars are Poisson. Interacting galaxies and those associated with substructure are not included in the observational fractions quoted. We also exclude the observed AGN and intense star-formers that (via visual inspection of Fig. \ref{Substructure}) may be part of substructure but haven't been identified via the selection criteria, and plot the curve for AGN and intense star-formers that we are confident are neither in substructure nor interacting.}
\label{FractionsAllV}
\end{center} 
\end{figure}

Fig. \ref{FractionsAllV} shows the AGN fraction as a function of clustercentric radius for the simulated AGN from our best-fitting ram pressure model, alongside the fraction of AGN and intense star-formers (both individually and combined) in the observed sample.
Note that the AGN fractions are normalised to the overall fraction of the sample over all radii and velocities to allow for a simple comparison. This figure shows that our model predicts a peak in the AGN fraction at lower clustercentric radii ($\sim$1$r_{\textrm{vir}}$) than the observed sample ($\sim$2$r_{\textrm{vir}}$). Modifying the model slightly would likely lead to a better match to the observations; for example, including slightly lower ram pressures would increase the number of simulated AGN at larger radii. However, the statistical significance of the discrepancy between the two samples is negligible; the model and observations are consistent within the errors, which are large due to the small sample size of the observations. 
Fig. \ref{FractionsAllV} also shows that the observed intense star-former fraction is found to decrease with clustercentric more rapidly than the observed AGN fraction. The differing radial distributions of the observed intense star-former and AGN samples suggests that AGN and star formation are subject to somewhat different mechanisms; a larger observational sample is needed to confirm this.

\begin{figure*}
\begin{center}
\includegraphics[scale=0.55]{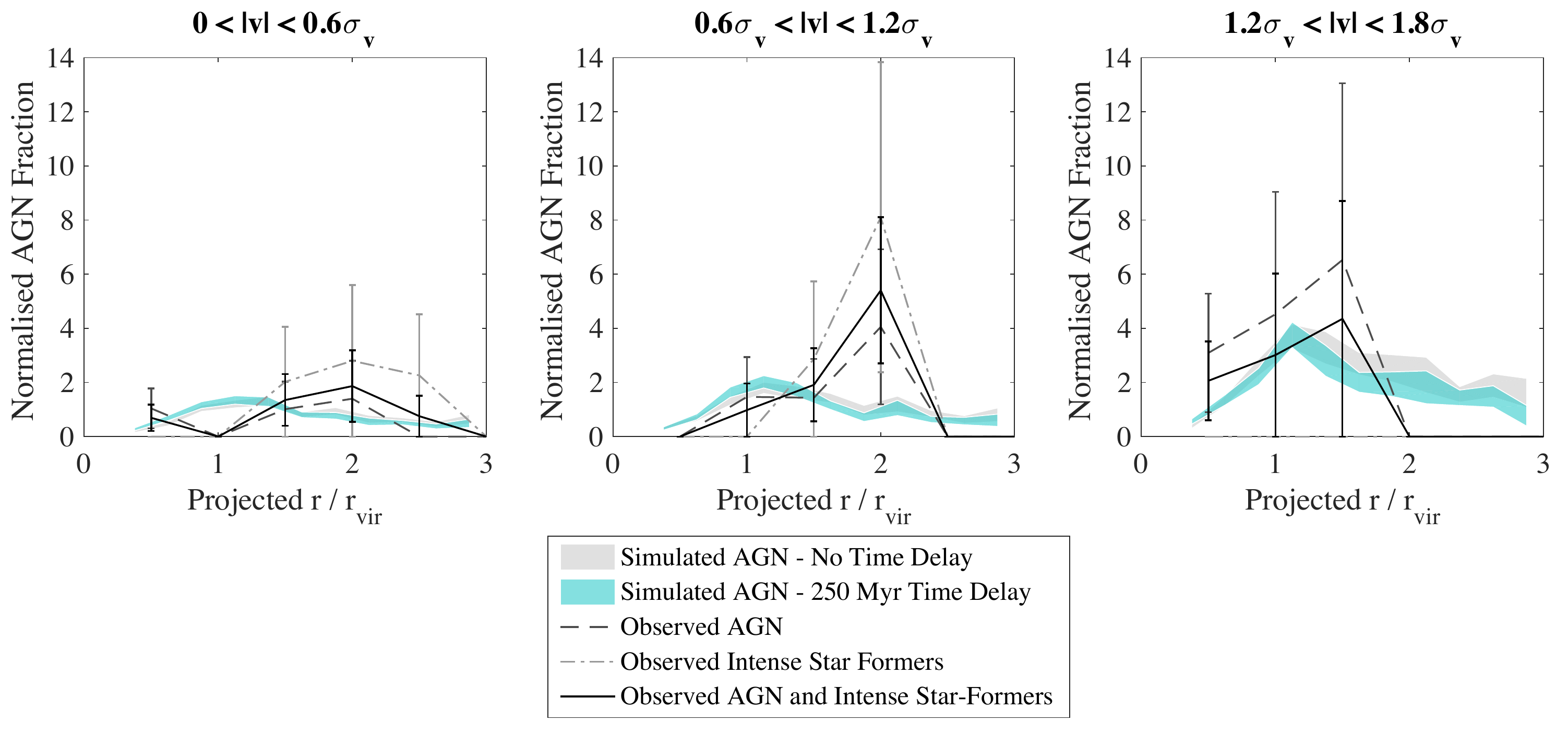} 
\caption{The AGN fraction as a function of clustercentric radius in velocity bins for the simulated AGN from the best ram pressure model (middle right panel of Fig. \ref{PimbbletPP}), alongside the fraction of AGN and intense star-formers (both individually and combined) in the observed sample, with the AGN fractions normalised to the overall fraction of the sample. 
The fraction of simulated AGN from models with no time delay and a 250 Myr time delay are both shown, with no significant differences found between the two. Radial bins for the observational sample are 0-0.75$r_{\textrm{vir}}$, 0.75-1.25$r_{\textrm{vir}}$, 1.25-1.75$r_{\textrm{vir}}$, 1.75-2.25$r_{\textrm{vir}}$, 2.25-2.75$r_{\textrm{vir}}$ and $>$2.75$r_{\textrm{vir}}$, whilst those for the simulated sample are finer: 0-0.5$r_{\textrm{vir}}$, 0.5-0.75$r_{\textrm{vir}}$, 0.75-1$r_{\textrm{vir}}$, 1-1.25$r_{\textrm{vir}}$, 1.25-1.5$r_{\textrm{vir}}$, 1.5-1.75$r_{\textrm{vir}}$, 1.75-2$r_{\textrm{vir}}$, 2-2.25$r_{\textrm{vir}}$, 2.25-2.5$r_{\textrm{vir}}$, 2.5-2.75$r_{\textrm{vir}}$ and $>$2.75$r_{\textrm{vir}}$. Error bars are Poisson. Interacting galaxies and those associated with substructure are not included in the observational fractions quoted. No intense star-formers from the observed sample have $1.2\sigma_v<v<1.8\sigma_v$.}
\label{Fractions}
\end{center} 
\end{figure*}

Fig. \ref{Fractions} also shows the AGN fraction as a function of clustercentric radius for the simulated AGN from our best-fitting ram pressure model and the observations, split into three velocity bins.
This figure shows that the ram pressure triggering prescription predicts that for $0<v<0.6\sigma_v$ and $0.6\sigma_v<v<1.2\sigma_v$ the AGN fraction peaks at lower radii than the observed peak ($\sim$1$r_{\textrm{vir}}$ compared with $\sim$2$r_{\textrm{vir}}$). For $1.2\sigma_v<v<1.8\sigma_v$, the ram pressure triggering prescription predicts a peak in the correct clustercentric radius range ($\sim$1--$2r_{\textrm{vir}}$). The most significant difference between the observations and model predictions is the peak in the AGN fraction for $0.6\sigma_v<v<1.2\sigma_v$. 
Three of the four observed AGN/intense star-formers in the peak bin are from the same cluster (A1620) and appear to be surrounded by galaxies with similar velocities (see Fig. \ref{Substructure}); these galaxies are not classified as substructure members due to the underdensity of galaxies in this region, but it is not unreasonable to suspect that these galaxies may be part of substructure. We visually inspect Fig. \ref{Substructure} to see if any additional AGN or intense star-formers appear to potentially be in substructure that were not identified by the substructure classification algorithm, and find one additional possibility in A1205. We exclude these 4 galaxies and plot the AGN fraction of those which we are confident are neither in substructure nor interacting in Fig. \ref{FractionsAllV}. Excluding these galaxies significantly reduces the observed AGN fraction around 2$r_{\textrm{vir}}$ in Fig. \ref{FractionsAllV}, bringing the model and observations into closer agreement; this would also have the same effect in the observed fraction in the $0.6\sigma_v<v<1.2\sigma_v$ range, reducing the AGN fraction in this bin to the point of being consistent with the model predictions.

\subsubsection{Time delay}
\label{sec:timedelayresults}
Investigations of different time delays show that any reasonable ($\leq 250$ Myr) delay between the onset of star formation and AGN activity has an insignificant effect on the observed position and velocity distributions of the AGN populations, with phase-space AGN distributions for all ram pressure triggering methods (Fig. \ref{PimbbletPP}) undergoing no significant changes when a time delay of 250 Myr is considered.
This can be clearly seen in Figs. \ref{FractionsAllV} and \ref{Fractions}, which show that under the best-fitting ram pressure model the AGN fraction as a function of clustercentric radius, both for the full velocity range and in velocity bins, is not significantly different for the cases with no time delay and a time delay of 250 Myr. 
A direct prediction of our model is that, since AGN and star formation are both expected to be triggered by the same mechanism, but with different time delays, we expect the AGN and star-forming galaxy distributions to be similar; that is, AGN and star-forming galaxies should be found at similar locations in clusters.

\subsubsection{AGN fraction}
Determining the AGN fraction produced by the model is not straightforward; the fraction of AGN is a convolution of the fraction of galaxies in which AGN are actually triggered by the ram pressure mechanism with the lifetime of AGN. Our model is unable to estimate the AGN duty cycle, and so assumptions must be made for this based on other studies.
The fraction of simulated galaxies which satisfy the best ram pressure triggering condition at the $z=0$ snapshot is 0.156, whilst the fraction of AGN in the observed sample is 0.010, excluding AGN undergoing mergers or in substructure.
Likely, not every galactic nucleus will be activated at observable luminosities, quite possibly depending on the detailed distributions of gas and stars. These AGN will then have a certain duty cycle.
It is beyond the scope of the present study to disentangle these processes, so we just give a lower limit for the duty cycle; for consistency between the observed and simulated samples, the duty cycle must be larger than 6 per cent. Whilst it is difficult to make comparisons to other studies with different mass selections and AGN triggering mechanisms, this estimate is consistent with the observations of \citet{Kauffmann04} and \citet{Ellison08}.

\section{General Model Predictions}
\label{sec:ModelPredictions}
\subsection{Model prediction for a complete sample of galaxies}
\begin{figure}
\begin{center}
\includegraphics[scale=0.5]{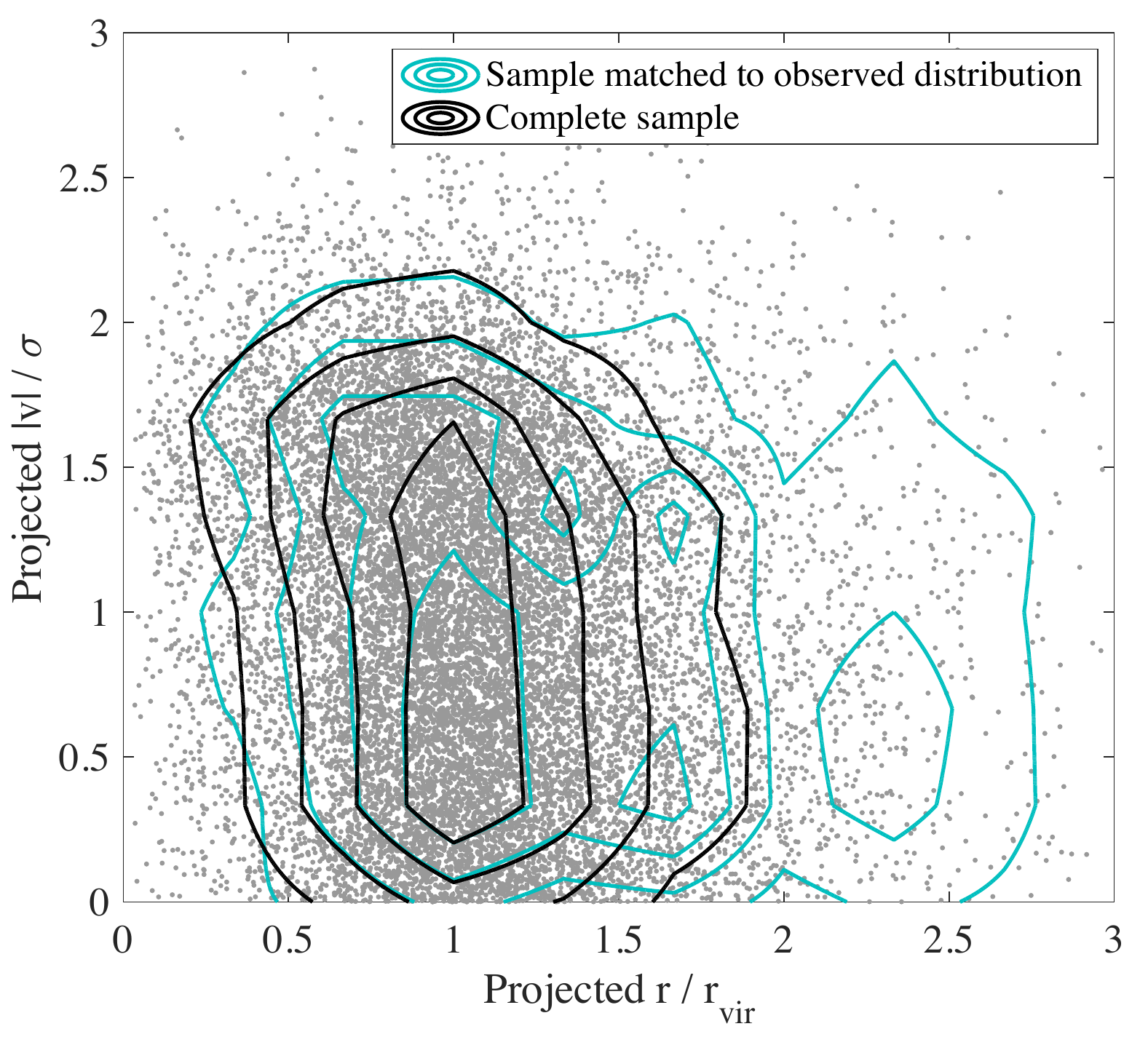} 
\caption{Phase-space diagrams of the simulated AGN produced by the best-fitting ram pressure method, for all galaxies with $\log(M_\ast/\textrm{M}_\odot)>10.4$. Black contours show the number density of simulated AGN at levels of 20, 40, 60 and 80\% of the peak number density, with the individual AGN shown as grey dots. Relative to the phase-space diagram for the simulated sample matched in radial and mass distributions to the \citet{Pimbblet} sample (middle right panel of Fig. \ref{PimbbletPP}; blue contours) the simulated AGN are shifted to lower clustercentric radii, with approximately unchanged (relative to Fig. \ref{PimbbletPP}) velocity distributions, due to the lack of selection effects against cluster core galaxies.}
\label{FullPP}
\end{center} 
\end{figure}

It is important to note that the distribution of AGN in the phase space as shown in Fig. \ref{PimbbletPP} results from galaxies that have a distribution in clustercentric radius and mass that matches that of the \citet{Pimbblet} sample. In reality, galaxies do not follow such a distribution: SAGE predicts many more galaxies at small clustercentric radii than those in the \citet{Pimbblet} sample.
This is most likely due to the requirement for targeted SDSS spectroscopy; that is, not all galaxies near the cluster centre are targeted by the observations. Fibre placement constraints make it difficult to observe all galaxies in dense regions such as cluster centres, with \citet{Yoon} estimating that the SDSS spectroscopic completeness can be as low as 65\% in the cores of rich clusters. Due to the size of the fibre plugs, the physical separation of SDSS fibre centres must be at least 55 arcseconds \citep{Blanton}, which corresponds to $\sim40$ kpc at the median redshift ($z=0.076$) of the \citet{Pimbblet} cluster sample.
A phase-space diagram of the AGN distribution for the simulated galaxies without this imposed distribution is given in Fig. \ref{FullPP}. This clearly shows a peak in AGN density at around the virial radius in projection, which is roughly uniformly spread across all projected velocities from 0-2$\sigma_v$.

\begin{figure}
\begin{center}
\includegraphics[scale=0.55]{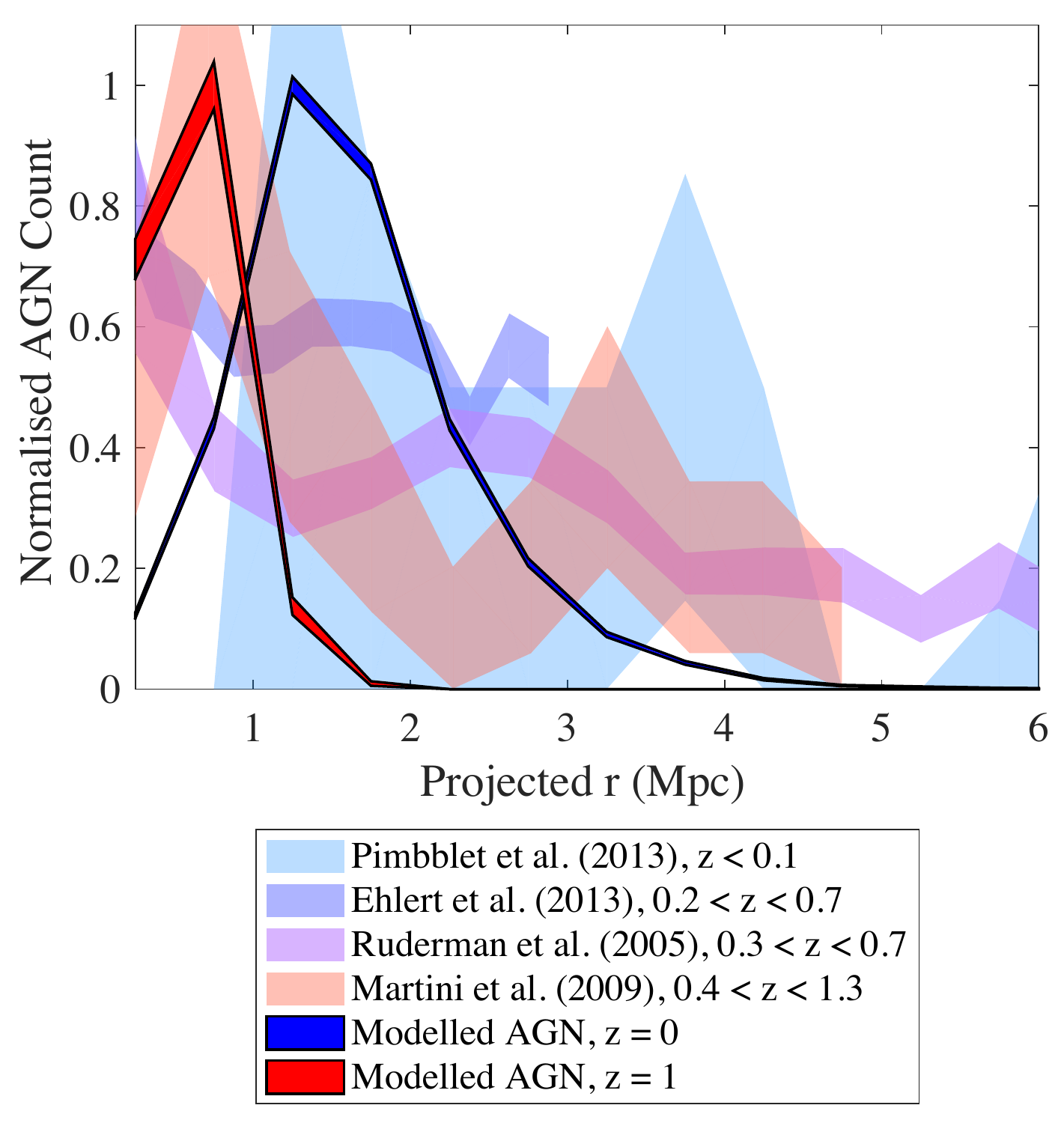} 
\caption{The number or number density of AGN in each observational sample as a function of radius. The peak of each distribution is normalized to unity. The normalised AGN count of the modelled AGN in the redshift 0 and 1 samples previously described are also shown.  }
\label{RadialFraction}
\end{center} 
\end{figure}

In their studies of $z\sim0.2$--0.7 X-ray cluster AGN, both \citet{Ruderman} and \citet{Ehlert1} found an AGN excess at approximately the viral radius. \citet{Ruderman} attributed this excess to an increased prevalence of gas-rich galaxy mergers that can induce AGN activity in this cluster-field transition region. The location of this excess is broadly consistent with the expectations of our ram pressure model, which also predicts an increased AGN prevalence at roughly the viral radius. We compare the radial distribution of X-ray AGN from these studies with our $z=0$ and 1 (see Section \ref{sec:highz}) model AGN samples in Fig. \ref{RadialFraction}. The X-ray AGN of \citet{Ehlert1} and \citet{Ruderman} show a slower decrease with radius, with a peak at the cluster centre and secondary peak at higher radii. The secondary peak in the \citet{Ehlert1} sample is consistent with the location of the peak in the ram-pressure triggered model AGN at $z=0$, whilst the $z=1$ model AGN peak at lower radii. The secondary peak in the \citet{Ruderman} sample is at slightly larger radii than the $z=0$ model AGN peak, at $r\simeq2.5$ Mpc compared with $r\simeq1.75$ for the model. Comprehensively comparing these datasets to our model is beyond the scope of this work, however this basic comparison with unmatched samples shows that the ram-pressure triggering model is at least broadly consistent with these observations.

\subsection{Cosmological evolution}
\label{sec:highz}
\begin{figure}
\begin{center}
\includegraphics[scale=0.55]{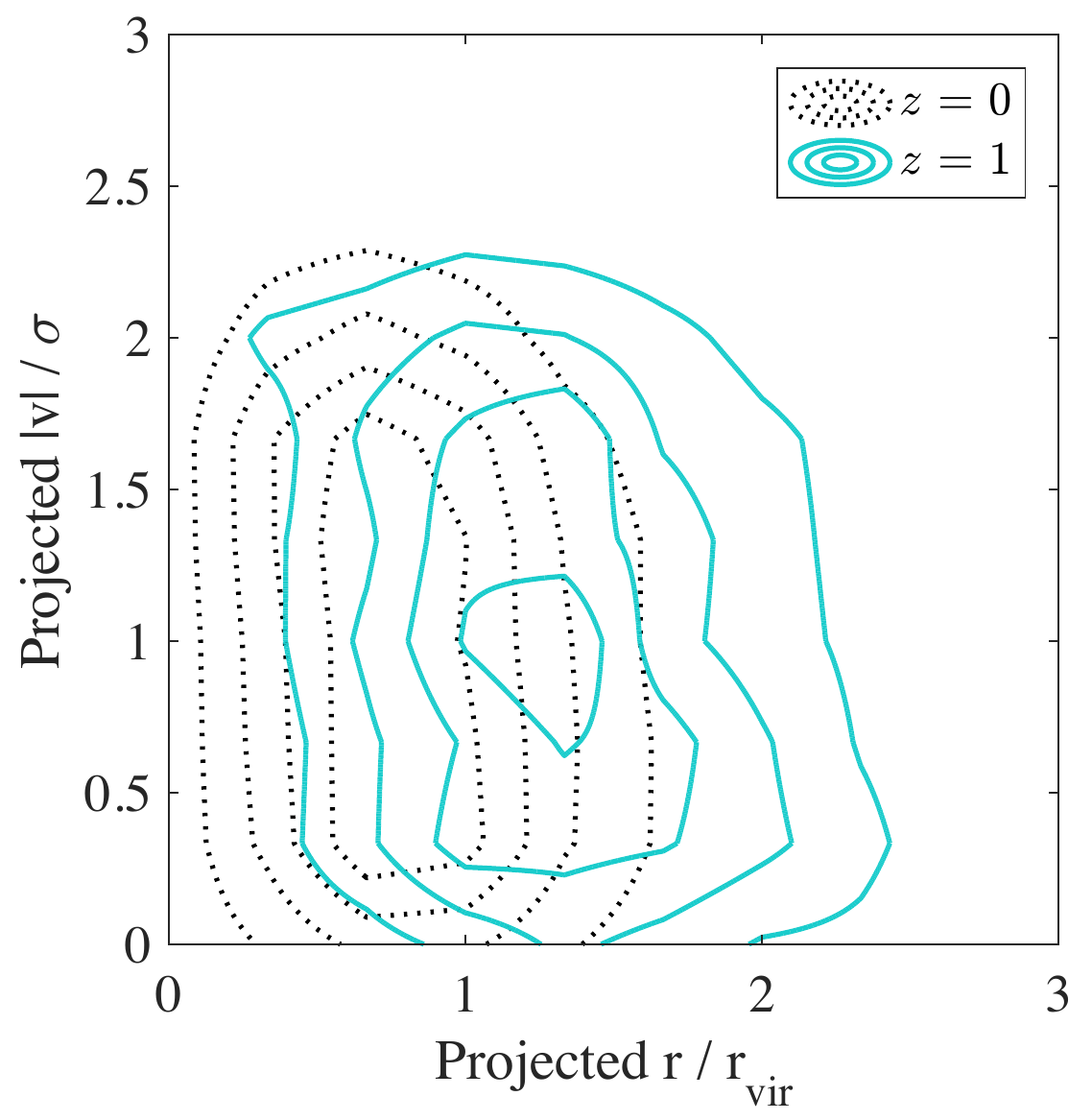} 
\caption{Phase-space diagrams of the simulated AGN for galaxy clusters at high ($z=1$; solid contours) and low redshift ($z=0$; dotted contours), with $M_{\textrm{vir}}>5\times10^{13}\textrm{M}_\odot$. Contours show the number density of simulated AGN at levels of 20, 40, 60 and 80\% of the peak number density.}
\label{Highz}
\end{center} 
\end{figure}

Our ram pressure triggering model can be applied to various samples of simulated galaxies to make predictions of the location of observed AGN or ram-pressure induced star-forming galaxies in such systems. For example, using additional snapshots at high redshift in SAGE, this model can also be applied to high redshift clusters. We consider galaxies with $M_\ast>10^{10.4}\textrm{M}_\odot$ in clusters with $M_{\textrm{vir}}>5\times10^{13}\textrm{M}_\odot$ at $z=1$, and show the resulting phase-space distribution of AGN in Fig. \ref{Highz}. This shows a shift of AGN to slightly larger clustercentric radii compared to $z\sim0$ to between roughly 1--2$r_{\textrm{vir}}$. This is likely due to virial radii being smaller at higher redshift for a given virial mass.
Note that pre-processing and galaxy-galaxy interactions will become more important at larger redshift, and so it is likely that the assumptions of relaxation and mergers being insignificant contributors to the observed AGN and intense star-former populations may become invalid (see Section \ref{sec:complications}).

Spectroscopic observations of X-ray counterparts in a $z\sim0.6$ cluster by \citet{Eastman} lead to the identification of 4 AGN with radii ranging from 0.51-2.35 $r_{\textrm{vir}}$. These values are consistent with those predicted by our model at higher redshift. \citet{Eastman} in addition considered clusters in the literature with comparable observations. AGN in the three clusters identified in $z\sim5.5$--6.4 lie predominantly at radii less than 1$r_{\textrm{vir}}$; these do not follow the expected distribution of our model, but individually are not at radii that are in conflict with the model results. It is also unclear whether this distribution is a result of selection effects. 

\citet{Martini09} considered X-ray AGN in massive clusters at $0.4 < z < 1.3$. These clusters are more massive than those in SAGE at $z=1$, which, alongside having no data for the general galaxy distribution in these clusters, makes a comprehensive phase-space comparison between the model and this dataset impossible. Instead, we plot the radial distribution of these AGN alongside our model AGN sample at $z=1$ in Fig. \ref{RadialFraction}. The model clearly reproduces the observed peak in the number of AGN at $r\simeq1$ Mpc; the \citet{Martini09} sample is in agreement with the ram-pressure triggering model. Note that Fig. \ref{RadialFraction} shows a tentative trend of the AGN count peaking at lower radii for higher redshift clusters.

In our model, the AGN candidate fraction is found to increase with redshift from 15.6 per cent at $z=0$ to 23.3 per cent at $z=1$. This is qualitatively consistent with studies
of AGN in clusters at redshifts up to $z\sim1.5$, such as \citet{Eastman} and \citet{Galametz}, which show that the cluster AGN fraction increases with redshift.\\

\subsection{Galaxy groups}
\begin{figure}
\begin{center}
\includegraphics[scale=0.55]{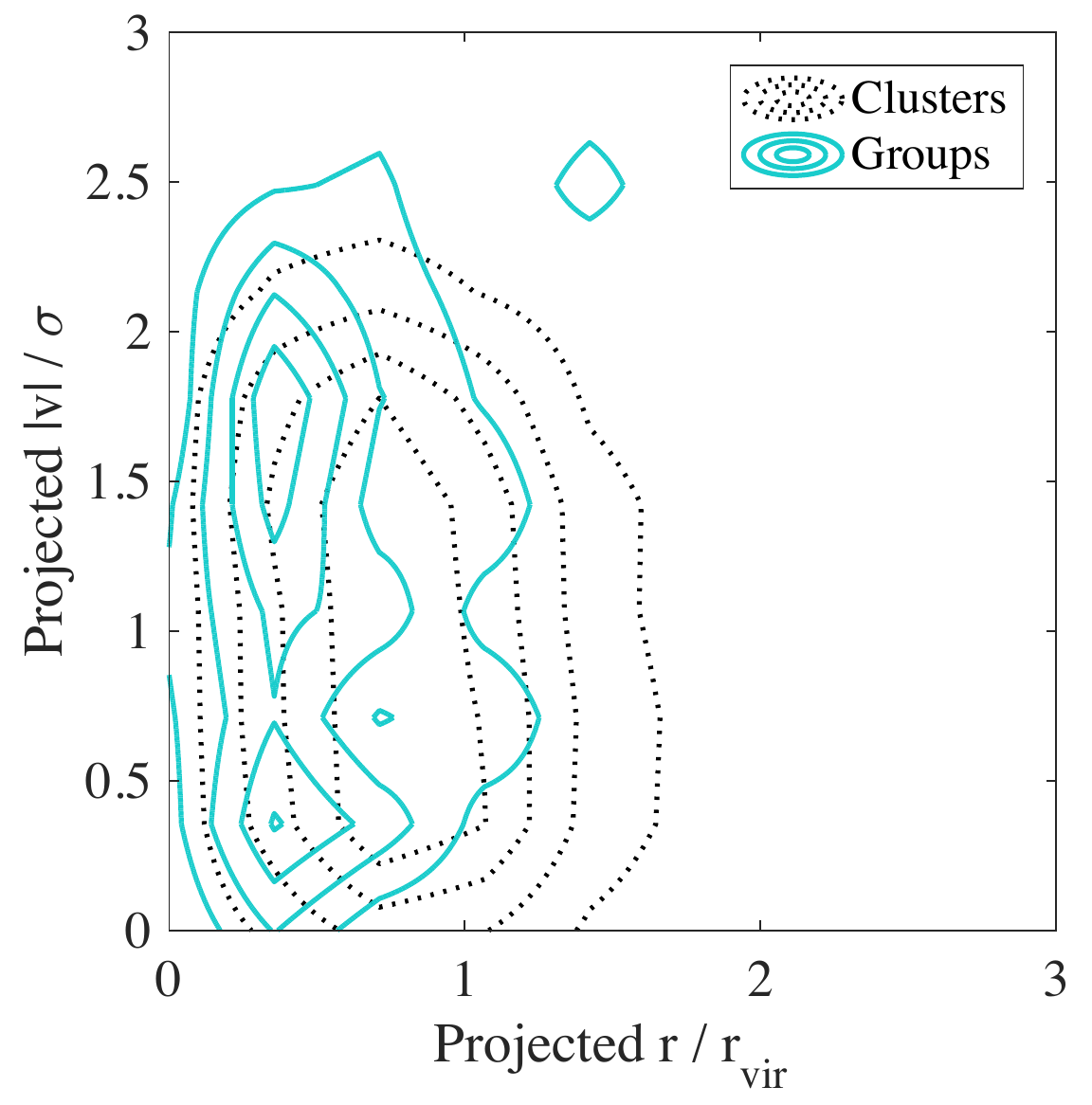} 
\caption{Phase-space diagrams of the simulated AGN for galaxy groups at $z=0$, with $0.5\times10^{13}\textrm{M}_\odot<M_{\textrm{vir}}<1.5\times10^{13}\textrm{M}_\odot$, and galaxy clusters, with $M_{\textrm{vir}}>5\times10^{13}\textrm{M}_\odot$. Contours show the number density of simulated AGN at levels of 20, 40, 60 and 80\% of the peak number density.}
\label{Groups}
\end{center} 
\end{figure}

To consider galaxy groups, we take SAGE galaxies with $M_\ast>10^{10.4}\textrm{M}_\odot$ in groups with $0.5\times10^{13}\textrm{M}_\odot<M_{\textrm{vir}}<1.5\times10^{13}\textrm{M}_\odot$, and show the resulting phase-space distribution of AGN in such systems in Fig. \ref{Groups}. This reveals a peak in expected AGN or star-forming activity at lower radii than for clusters, with most AGN having $0<r<1r_{\textrm{vir}}$. Since the gas density in groups is lower at a given distance from the centre than that of clusters, the ram pressure group galaxies experience at a given velocity will be lower, and so to lie in the ram pressure triggering range galaxies must lie closer to the group centre. We note that the density profile adopted in Fig. \ref{Groups} comes from the average of cluster observations by \citet{Vikhlinin}, and hence may not be strictly applicable.

\begin{figure}
\begin{center}
\includegraphics[scale=0.55]{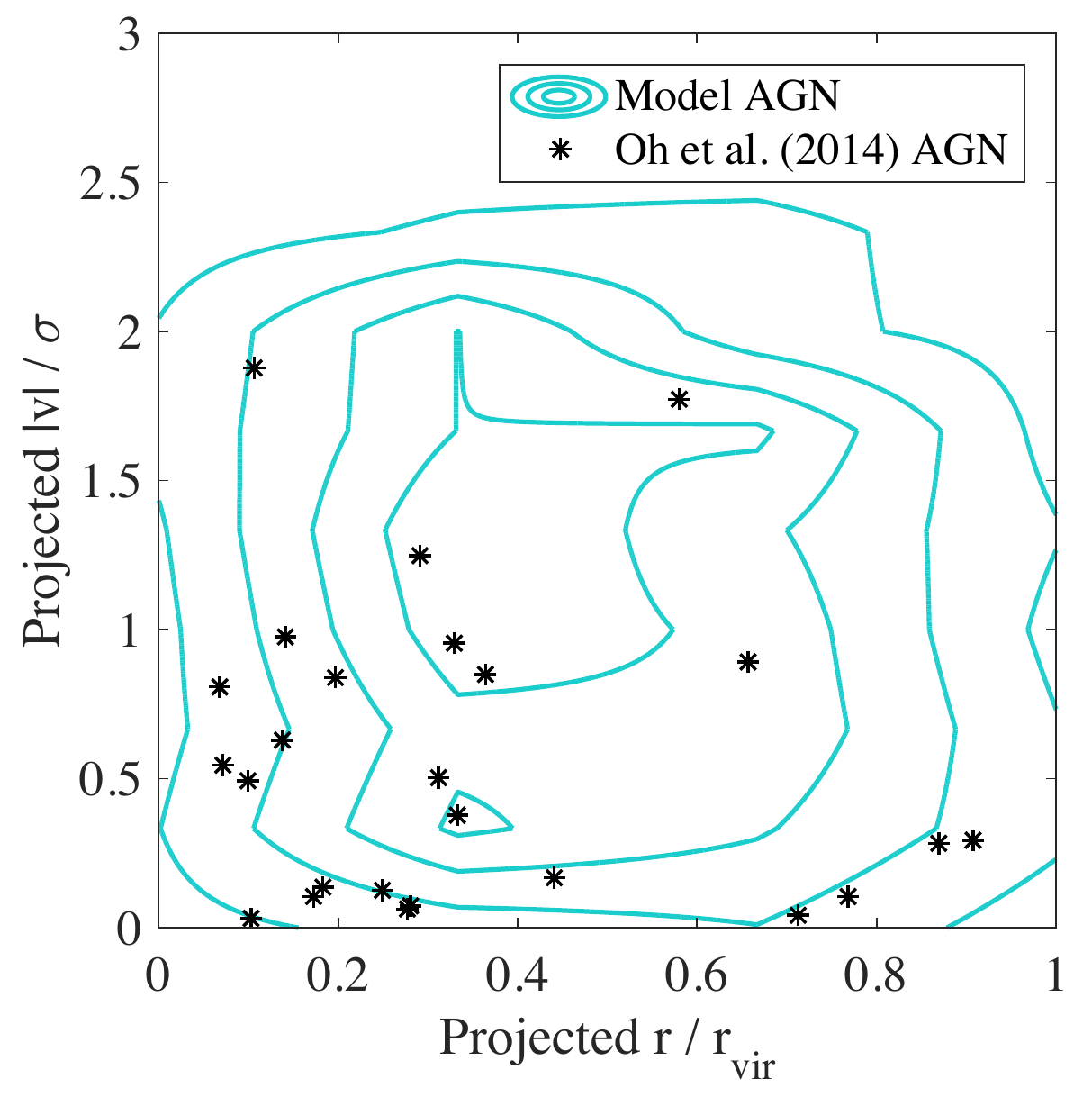} 
\caption{Phase-space diagrams of the simulated AGN for galaxies with $M_\ast>10^{10.4}\textrm{M}_\odot$ in groups with $12.7<\log M_{\textrm{vir}}/\textrm{M}_\odot<14.7$ at $z=1$, and with radial distribution matched to the distribution of general cluster galaxies in the \citet{Oh} sample. AGN from the \citet{Oh} observational sample are also plotted. Contours show the number density of simulated AGN at levels of 20, 40, 60 and 80\% of the peak number density.}
\label{Oh}
\end{center} 
\end{figure}

\citet{Oh} considered X-ray AGN in galaxy groups at $0.5<z<1.1$, and found them at lower radii ($\lesssim 0.4 r_{\textrm{vir}}$) than the general galaxy population. In order to compare this data set with our model predictions, we take galaxies in clusters of the same mass range ($12.7<\log M_{\textrm{vir}}/\textrm{M}_\odot<14.5$) in the $z=1$ snapshot. We then match the radial distribution of the simulated sample to that of the observed group galaxies, which have radii less than $r_{\textrm{vir}}$. We implement a mass cut of $M_\ast>10^{10.4}\textrm{M}_\odot$ to the simulated galaxies as in the \citet{Pimbblet} sample, since stellar masses and thus a mass cut are not measured by \citet{Oh}. The resulting AGN phase space distribution for the matched model galaxies is compared with the observed \citet{Oh} sample distribution in Fig. \ref{Oh}.
It can be seen that the observed AGN lie within the expected regions of phase-space by the ram pressure triggering model, with a slightly lower spread in velocities than predicted by the model; the ram pressure model is in reasonable agreement with the observed sample of groups at high redshift, providing further validation for the model. We note that the distribution of these high-redshift group AGN are not well described by either the low-redshift group sample or the high-redshift cluster samples shown in Figs. \ref{Highz} and \ref{Groups}.

\section{Discussion}
\label{sec:discussion}
\subsection{Model predictions and implications}
The model of ram pressure triggering where galaxies are triggered as AGN or star-forming galaxies if $2.5\times10^{-14}<P_{\textrm{ram}}<2.5\times10^{-13}$ Pa and $P_{\textrm{ram}}/P_{\textrm{internal}}>2$ provides a good explanation of the observed distribution of AGN and intense star-formers in the radius--velocity phase space. This suggests that ram pressure might indeed act to compress the gas in a galaxy, leading to an increase in star formation and potentially inducing AGN activity. 

Alternatively, hydrodynamical simulations show that if the ram pressure is too large, the gas will be stripped from the galaxy. By considering the location of simulated SAGE cluster galaxies with high ram pressures ($2.5\times10^{-13}<P_{\textrm{ram}}<2.5\times10^{-12}$ Pa), such ram pressure stripping is expected to occur for galaxies at low clustercentric radii ($<1r_{\textrm{vir}}$; see Fig. \ref{PimbbletPP}); galaxies in the central regions of clusters are likely to undergo ram pressure stripping. This is qualitatively consistent with observational studies such as \citet{Wetzel} which find that the fraction of quenched satellite galaxies increases with decreasing clustercentric radius; the  \citet{Luo} ram pressure stripping model which strips gas from the outer radii of satellite galaxies where $P_{\textrm{ram}}>P_{\textrm{internal}}$ also produces this signature.

\subsection{Complications and caveats}
\label{sec:complications}
Several caveats are associated with our analysis. 
Firstly, individual galaxy interactions are expected to (at least sometimes) trigger AGN, in addition to the ram pressure which acts on all cluster galaxies;
galaxy interactions and mergers are commonly associated with AGN triggering, with the AGN fraction of pair galaxies found to increase with decreasing galaxy separation \citep[e.g.][]{Ellison,Woods}. The AGN fraction of pair galaxies is found to increase by up to a factor of 2.5 for pairs with projected separations of less than 40$h^{-1}$ kpc \citep{Ellison}. 
However, less than 1\% of the simulated SAGE galaxies in our low-redshift sample at a given time have a neighbour within 40$h^{-1}$ kpc and so are undergoing a merger that may trigger AGN activity. In comparison, over 10\% of galaxies at a given time have a ram pressure and pressure ratio in the best-fitting triggering parameter range, as detailed in Section \ref{sec:BestModel}. Hence, too few mergers occur for this to be the dominant triggering mechanism for AGN in low-redshift clusters. This additional triggering mechanism can be ignored in our low-redshift simulations; however, this may not be the case at higher redshift where galaxy interactions are more common.


It is also important to note that our triggering models are independent of the galaxy's gas properties. Galaxies triggered as AGN by our models are not required to have a given amount of gas; since an abundant supply of cold gas is necessary to fuel black holes and power AGN, a triggering condition that depends on the gas properties of each galaxy may be more appropriate. The internal pressure assumed for each galaxy takes no account of the galaxy's gas content, whilst the \citet{Blitz} pressure prescription does; if this were instead used, the triggering prescription would be more dependent on the gas properties of each galaxy. However, the \citet{Blitz} prescription requires detailed knowledge of the gas content, including velocity dispersions and radial distributions, and so would be difficult to implement without detailed hydrodynamical simulations. In addition, AGN properties such as luminosity may depend on the gas content of a galaxy; this could lead to an altered luminosity distribution of the simulated AGN.

Uncertainties arise in the temperature and density profiles assumed for each cluster, which leads to uncertainties in the ram pressure and internal pressure calculated for each galaxy. These uncertainties are expected to be less than the order of magnitude variations considered in the triggering ram pressures, however, and thus should not affect our results significantly. Uncertainties in temperature and density are expected to increase with radius, because at larger distances from the cluster centre they are harder to measure and substructure plays more of a role; this may cause our results to be less accurate at larger radii ($\gtrsim 2 r_{\textrm{vir}}$). Since the density profiles considered here are both smooth and spherically symmetric, future work should consider density profiles that are more physically reasonable; for example, density profiles from hydrodynamical cosmological simulations would allow for a clumpy ICM, more representative of unrelaxed clusters.

An important feature of the \citet{Pimbblet} observations is that they are of low-redshift relaxed clusters; while this ram pressure triggering model reasonably predicts the AGN distribution of this sample, this is a very specific subset of the general cluster population. Unrelaxed clusters are those undergoing interactions with other clusters or subclusters. The galaxies in the cluster or subcluster can be pre-processed by their environment prior to their accretion into the primary cluster, due to gravitational interactions, mergers and ram pressure stripping \citep[e.g.][]{FujitaPP,Vijay,Cybulski}. These pre-processed AGN complicate studies of the effects of the general cluster environment on AGN activity, hence the \citet{Pimbblet} sample of relaxed clusters is an ideal sample for determining the effects of the environment on AGN activity.

A final notable caveat is that filaments are observed in some of the \citet{Pimbblet} clusters (see Fig. \ref{Substructure}). AGN may be caused by pre-processing of groups in these filaments before they are accreted onto the cluster \citep[see e.g.][]{Porter}. This may cause AGN to lie preferentially along filaments, and although these should be identified by our substructure detection method, this could lead to an overestimate in the true amount of AGN caused by the general environment at larger radii where such filaments are found. In addition, few interloping galaxies/AGN may be present in the observations \citep[see e.g.][]{Pimbblet2011}, which may have a minor effect on the results.

\section{Conclusions}
\label{sec:conclusions}
Hydrodynamical simulations of the effect of ram pressure on gas-rich galaxies suggest that below the regime of ram pressure stripping the enhanced pressure might lead to an elevated level of star formation and the onset of AGN activity. We have tested this effect with a semi-analytic galaxy evolution model based on the Millennium simulation and compared it to an observational sample of galaxies in low-redshift clusters. The phase-space properties of observed AGN populations are consistent with a triggering scenario for intermediate ram pressures. The critical range corresponds to ram pressures expected around the virial radius in low-redshift galaxy clusters, and agrees with expectations from detailed hydrodynamical simulations. If AGN were triggered preferentially at high ram pressures, such as those considered relevant for ram pressure stripping, the model would predict an AGN population at significantly lower clustercentric radii than observed. We make predictions for high-redshift clusters and poor groups of galaxies, which are broadly consistent with observations.

Our analytical model is complementary to detailed cosmological hydrodynamic simulations \citep[e.g.][]{Vogelsberger,Crain,Kaviraj,Dave}, and may assist with interpretation of current and future observations, including with deep multi-wavelength surveys of groups and clusters such as the Galaxy and Mass Assembly (GAMA) project \citep{Driver} and its successor 4MOST WAVES; and integral field surveys including SAMI \citep{Croom}, CALIFA \citep{Sanchez} and MANGA \citep{Bundy}.

Our model for ram pressure triggering of AGN is complementary to new spectroscopic observations of 7 galaxies which show signs of ram pressure stripping \citep{Poggianti}. These show that galaxies with ram-pressure stripped gas tails are highly likely to host an AGN, with 6 of the 7 observed galaxies showing AGN signatures. Whilst this is a very small sample, it therefore seems likely that ram pressure can indeed trigger AGN activity, as predicted by our work.

\section*{Acknowledgements}

We thank the referee for their constructive comments that helped to improve the paper, and Andrew Cole and Simon Ellingsen for feedback on an earlier version of this work. MAM thanks the University of Tasmania Foundation for Honours scholarships, and the faculty of Science, Engineering and Technology for a summer research scholarship. SSS and MGHK thank the Australian Research Council for an Early Career Fellowship (DE130101399).




\bibliography{TriggeringAGN_Marshall} 







\bsp	
\label{lastpage}
\end{document}